\documentclass[aps,pra,twocolumn,10pt,eqsecnum]{revtex4-1}
\usepackage{amsfonts,amssymb,amsthm,amsmath, mathdots,bm,esint}
\usepackage{varwidth}
\usepackage{dsfont}
\usepackage{mathrsfs}
\usepackage{graphicx}
\usepackage{tcolorbox}
\usepackage{hyperref}
\usepackage{xcolor}
\usepackage{physics}
\usepackage{multirow}
\usepackage{hieroglf}
\usepackage{units}
\usepackage{todonotes}
\usepackage{algpseudocode}
\usepackage{algorithm}
\usepackage{algorithmicx}
\algrenewcommand\algorithmicrequire{\textbf{Input:}}
\algrenewcommand\algorithmicensure{\textbf{Output:}}
\usepackage[]{qcircuit}
\usepackage{pgf}
\tcbset{colback=white,center title}
\newcommand{\ground}{
\begin{pgfpicture}{0cm}{0cm}{.5cm}{-.2cm}
\pgfline{\pgfxy(.25,.05)}{\pgfxy(.25,-.15)}
\pgfline{\pgfxy(.05,-.15)}{\pgfxy(.45,-.15)}
\pgfline{\pgfxy(.1,-.2)}{\pgfxy(.4,-.2)}
\pgfline{\pgfxy(.15,-.25)}{\pgfxy(.35,-.25)}
\end{pgfpicture}}

\newtheorem{cert}{Certificate}[]

\theoremstyle{definition}

\usepackage{commath}
\newcommand{\missingcite}{{\color{red} [~]}}
\begin{document}
\title{Adversarial quantum teleportation}
\author{Nehad AttaElmanan AbdElrahim Mabrouk\href{https://orcid.org/0000-0001-5275-3475}{\includegraphics[scale=0.05]{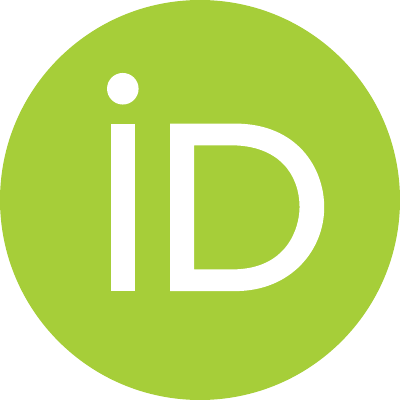}}}
\email{nehad.attaelmananabd@ucalgary.ca}
\affiliation{Institute for Quantum Science and Technology, University of Calgary, Alberta T2N 1N4, Canada}
\author{Barry C.\ Sanders\href{https://orcid.org/0000-0002-8326-8912}{\includegraphics[scale=0.05]{figures/orcidid.pdf}}}
\email{sandersb@ucalgary.ca}
\affiliation{Institute for Quantum Science and Technology, University of Calgary, Alberta T2N 1N4, Canada}
\begin{abstract}
Claims of successful quantum teleportation are backed up by showing that fidelity exceeds some specified threshold,
but whether fidelity is the performance metric and what the threshold should be has been a subject of vigorous debate.
We construct adversarial models for quantum teleportation,
i.e., involving cheating parties,
and show that fidelity thresholds can be justified in the context of the type of adversary trying to prove unsuccessful quantum teleportation has been successful.
In particular we show how previously established average-fidelity thresholds of $\nicefrac12$ and $\nicefrac23$ arise from our adversarial approach.
Mathematically,
we describe adversarial quantum teleportation as a multi-partite protocol with explicit quantum-logic circuits in both honest and cheating settings,
and our methods are relevant beyond quantum teleportation to other quantum-information gadgets.
\end{abstract} 
\date{\today} 
\maketitle
 \section{Introduction}
\label{sec:introduction}
Quantum teleportation is one of the earliest and most important quantum information tasks,
and the purpose of quantum teleportation is to send quantum information through a classical channel by exploiting the prior shared entanglement and classical communication~\cite{BBC+93}
with the prior entanglement necessary for teleportation to be `non-local~\cite{GBZ+16}.
Quantum teleportation is valuable for quantum communication~\cite{PhysRevLett.81.5932,ZZHE93}
and quantum computation via teleporting gates~\cite{Gottesman1999},
but, in the real world, 
quantum teleportation is never perfect~\cite{PEW+15,HGL+23}.
Due to the imperfect nature of quantum teleportation,
an experimental demonstration of quantum teleportation is deemed to be successful if its performance benchmark exceeds some agreed threshold.
Historically,
quantum teleportation has been characterized by fidelity~$f$,
which is the average of each fidelity $f$~\cite{BFK01}
over all input states according to some agreed prior of input states,
such as uniform or Gaussian~\cite{braunsteinK98}.

Typically, some threshold fidelity~$f_\text{\text{th}}$
is introduced such that $f>f_\text{\text{th}}$ proves teleportation is `quantum',
e.g.,
for
\begin{equation}
\label{eq:fthhalftwothirds}
f_\text{\text{th}}=\nicefrac12,\,
f_\text{\text{th}}=\nicefrac23,
\end{equation}
according to conditions proposed by Braunstein and Kimble~\cite{BK98}
and by Grosshans and Grangier~\cite{GG01},
respectively.
Here we introduce a framework for establishing~$f_\text{\text{th}}$ based on a general adversarial (i.e., cheating) model,
and we show exactly how $f_\text{\text{th}}=\nicefrac12$
and $f_\text{\text{th}}=\nicefrac23$
arise from our model.
Our analysis explains clearly quantum-teleportation fidelity thresholds and sets the stage for analyzing quantum logic gates~\cite{BBC+95}
by adversarial principles.
We refer to the process of cheating to pretend quantum teleportation has been achieved as `spoofing' here.

We build our framework by representing quantum teleportation employing quantum-circuit diagrams and modifying the quantum circuit to represent various cheating strategies.
The original quantum teleportation protocol (QTP),
which we denote~$\wp$
(and use~$\wp$ with qualifiers for each of the different QTPs introduced here)
involves sender~A and receiver~B.

To generalise QTPs,
we add two agents to construct the QTP~$\wp^0$:
C, who generates the input and measures the output,
and~D, who supplies entanglement.
C's task is to verify quantum teleportation,
and,
as~A and~B are not connected by a quantum channel,
D~supplies entanglement in advance of a given experimental run.
In our analysis,~C creates a multipartite quantum state and requests~A to teleport one share to~B;
this QTP
reduces to~C sending the entire state to~A for teleportation to~B
if the multipartite state is the simplest case of a unipartite state.
Our cheating strategies correspond to
(i)~A measuring and choosing to send this information to honest~B (or not in some instances),
(ii)~honest~A sending to dishonest~B
(who measures),
and
(iii)~both~A and~B dishonest.

Our paper is organized as follows.
We provide a salient background on quantum teleportation, mathematical formalism, and quantum-circuit diagrams in~\S\ref{sec:background}.
Then in~\S\ref{sec:approach},
we describe our model comprising four agents of which two (A~and~B) could cheat
followed by our mathematical analysis.
Finally, we present our methods for modifying the circuit for cheating.
In particular, we present our results for fidelity in different cases in~\S\ref{sec:results}
and discuss these results in~\S\ref{sec:discussion}.
In~\S\ref{sec:conclusions}
we provide an overview of our results and conclude with an outlook for future work.
\section{Background}
\label{sec:background}

In this section, we present the essential background required for subsequent analysis.
First, in~\S\ref{subsec:qteleportation},
we review key quantum-teleportation results including the concept, extensions and state-of-the-art theory and experiments.
Then
in~\S\ref{subsec:fidelity},
we review key mathematical constructs concerning state fidelity measures.
Finally,
in~\S\ref{subsec:qcircuitdiagram},
we describe how quantum-circuit diagrams represent quantum teleportation.

\subsection{Quantum teleportation}
\label{subsec:qteleportation}
We now summarize key elements of quantum teleportation beginning with a discussion of ideal quantum teleportation,
meaning that the QTP is described only in terms of pure states, unitary channels, perfect measurements and perfect communication.
Then we elaborate on the mathematical description for the ideal quantum teleportation.
Finally, we summarize state-of-the-art results pertinent to our analysis.

Quantum teleportation is a two-party QTP, for sending quantum information from one party to another,
in the absence of a quantum channel,
by instead sharing prior entanglement and transmitting classical information.
As the term `protocol' is sometimes used loosely in the physics community,
we clarify that we are using the term carefully in its computer science meaning,
namely rules for transmitting and receiving data between nodes of a network.
In the original QTP proposal,
this procedure is described by
``cleanly separat[ing the state]
into classical and quantum parts''~\cite{BBC+93}.
The simplest case concerns a transmitting agent~A teleporting a qubit,
represented by a vector in two-dimensional projective Hilbert space,
namely,
\begin{equation}
\label{eq:qubit}
\ket{\psi}
    =\psi_0\ket0+\psi_1\ket1
    \in\mathscr{H}_2\cong\mathbf{CP}^2
    \ni\psi=\begin{pmatrix}\psi_0\\ \psi_1\end{pmatrix},
\end{equation}
with $\abs{\psi_0}^2+\abs{\psi_1}^2=1$
and~$\mathscr{H}_2$
denoting two-dimensional Hilbert space
spanned by the two orthonormal `logical states'~$\ket0$ and~$\ket1$.

General Hilbert space is denoted~$\mathscr{H}$
with dimension~$\aleph_0$ (countably infinite).
The last symbol in Eq.~(\ref{eq:qubit}) denotes the two-dimensional complex projective space.
We employ~$\psi$ 
as the widely adopted label for general quantum states.
This qubit~(\ref{eq:qubit}) is sent to receiving agent~B,
although note that the original QTP paper~\cite{BBC+93}
includes analysis of more general qudits~\cite{WHSK20}.

A and~B are each supplied in advance
with one qubit share from a two-qubit maximally entangled state (ebit)~\cite{BBC+93}.
We simplify tensor-product notation by replacing~$\mathscr{H}^{\otimes m}$
by~$\mathscr{H}^m$,
so the ebit is
\begin{equation}
\label{eq:ebit}
\ket{\widetilde{\bm\epsilon}}
:=\ket{0\;\varepsilon_1}
+(-1)^{\varepsilon_0}\ket{1\;1-\varepsilon_1}
    \in\mathscr{H}_2^2,\,
\bm\epsilon:=\left(\varepsilon_0,\varepsilon_1\right)
\end{equation}
and $\ket{\pm}:=(\ket0\pm\ket1)/\sqrt2$.
Generalising Eq.~(\ref{eq:qubit}) to the two-qubit case, we write
\begin{equation}
\label{eq:twoqubits}
\ket{\Psi}
    =\sum_{i,j=0}^1\psi_{ij}\ket{ij}, \,
    \psi=\begin{pmatrix}\psi_{00}\\
\psi_{01}\\ \psi_{10}\\
    \psi_{11}\end{pmatrix}\in\mathbf{CP}^4
\end{equation}
for~$\mathbf{CP}^4$
four-dimensional complex projective space.

Linear operators,
which are homomorphisms on~$\mathscr{H}$,
are `lifted' from states by using the inner product to map a vector~$\ket\psi$ to a covector~$\bra\psi$,
and~$\ket\psi\bra\psi'$
is an operator,
which reduces to a projector if~$\psi'=\psi$.
Unitary operators are an isometric subset of linear operators.
Mathematically,
a generic linear operator is~$\Theta$
such that
\begin{equation}
\label{eq:liop}
\mathcal{L}(\mathscr{H})
\ni\Theta:\mathscr{H}\to\mathscr{H}:\ket\psi\mapsto\Theta\ket\psi,
\end{equation}
and a unitary operators satisfies
$\Theta^\dagger=\Theta^{-1}$.
We denote unitary operators by~$\mathcal{U}(\mathscr{H})$.

For $d$-dimensional Hilbert space~$\mathscr{H}_d$,
a unitary operator can be represented as an element of the Lie group SU$(d)$
with smallest faithful representation being a $d\times d$ complex matrix~\cite{Wyb74}.
We follow the standard abuse of notation in theoretical physics of equating the operator with its representation.
A mixed state is a positive-definite trace-class self-adjoint linear operator:
\begin{equation}
\label{eq:selfadjop}
\mathcal{B}(\mathscr{H})\ni\rho:\mathscr{H}\to\mathscr{H}
\end{equation}
such that
$\rho^2=\rho\Leftrightarrow\rho$ is pure $\Rightarrow\rho=\ket\psi\bra\psi$.
We can thus write~$\rho$ as a $d\times d$ Hermitian matrix.

\begin{figure}
  \begin{tcolorbox}
   \[
\begin{array}{c}
\begin{minipage}{\columnwidth}
\Qcircuit @C=1.2em @R=1.5em {
&(a)&&&(b)&&&(c)&&&&\\
 & \gate{R(\psi)}&\qw&&\meter&\cw&& \ctrl{1}&\qw&&& \\  & \gate{H}&\qw&&\ground \qw&&&\targ&\qw}
\end{minipage}
\\[1cm]
\begin{minipage}{\columnwidth}
\Qcircuit @C=1.2em @R=1.5em {
&&(d)&&&&(e)&&&\\
   &{/^m}\qw&\gate{R(\Psi)}&{/^m}\qw&\qw&&&&\gate{G}\cw&\cw
 }
\end{minipage}
\end{array}\]
\end{tcolorbox}
\caption{Elements of quantum circuits.
(a)~Single-qubit rotation gate (top) and Hadamard gate (bottom),
(b)~measurement and bit feedback (vertical double line)(top) and ignoring a qubit (tracing out a qubit)(bottom),(c)~CNOT gate, (d) a wire for $m$-qubit with $m$-qubit rotation gate,~(e) random-bit generator with the input implicitly being logical 0.%
}
\label{fig:operations}
\end{figure}
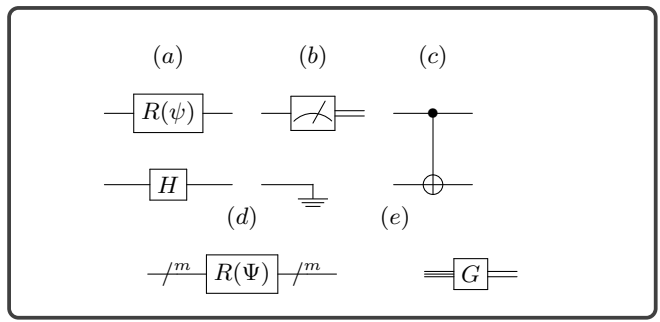

The Bell-state measurement (BSM) is a key part of the QTP.
A~performs a BSM on her single-qubit share of~$\ket{\widetilde{00}}$~(\ref{eq:ebit})
along with her to-be teleported qubit
and transmits to~B a two-bit message conveying which of four BSM outcomes occurred.
Finally,~B applies one of four single-qubit Pauli gates,
shown in Fig.~\ref{fig:operations},
to~B's share of~$\ket{\widetilde{00}}$,
thereby converting B's share to~A's original share of~$\ket\Psi$.

{The original QTP~$\wp$
is shown in Fig.~\ref{fig:qteleportation} and is a two-party protocol involving only agents~A and~B.
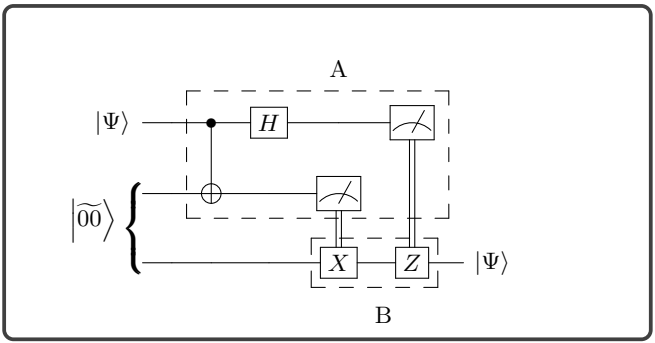
\begin{figure}
\begin{tcolorbox}
\[
\begin{array}{c}
\Qcircuit @C=1.2em @R=1.5em {
& &  & &\mbox{A}\\
\lstick{\ket\Psi} & \qw& \ctrl{1} &\gate{H}&\qw & \meter& \\ &  \qw & \targ & \qw&\meter&&\\  & \qw & \qw&\qw&\gate{X}\cwx & \gate{Z}\cwx[-2] &\qw&&\lstick{\ket\Psi}\\
& ~~~~~~~~~~&~~~ &~~~&~~~~~~~~~~~\mbox{B}
\gategroup{2}{3}{3}{6}{1.2em}{--}\gategroup{4}{5}{4}{6}{.75em}{--}
\gategroup{3}{1}{4}{1}{.8em}{\{}
\inputgroupv{3}{4}{0.7em}{1.1em}{\ket{\widetilde{00}}~~} }
\end{array}\]
\end{tcolorbox}
\caption{%
Quantum circuits representing standard single-qubit teleportation~$\ket{\Psi}$ with resource~$\ket{\widetilde{00}}$
and input~$\ket{\Psi \widetilde{00}}$.
A~applies the entanglement-gadget inverse~(\ref{eq:inverseentgadget}),
then measures her qubits,
and communicates the measurement outcome $(a,b)$ to~B. The vertical double line~$\|$ represents classical communication.
Finally, B~executes X, Z, or XZ operations depending on the outcomes of A's measurements.
A (pure) qubit is represented as a horizontal line, or `wire',
flowing to the right through the gates shown in Fig.~\ref{fig:operations}.}
\label{fig:qteleportation}
\end{figure}
The unknown state~$\ket{\Psi}$ is provided to~A along with a share of the entangled state~$\ket{\widetilde{00}}$, and the other share is provided to~B.
A~performs a joint measurement and communicates the results to~B,
who then uses this information to reconstruct the original quantum state.

Although quantum teleportation is conveniently described in terms of qubits, the concept of quantum teleportation extends to other models including qudits ($d$-dimensional system and $d=2$ for a qubit)~\cite{BBC+93}, continuous-variable quantum information~\cite{PhysRevA.49.1473,braunsteinK98},
non-maximally entangled states~\cite{MSM95},
mixed states~\cite{adhikari2016teleportation}
and concatenated QTPs~\cite{BS05}.
All these QTPs have in common an entanglement resource, a share of a multipartite state to be teleported, a BSM or a higher-dimensional extension and a share of the initial state, classical communication from the sender to receiver, a local unitary transformation, and verification, typically based on average gate fidelity
(and extendsions such as fidelity deviation~\cite{BRK18}).

Experimental implementations of QTP have been claimed over a variety of physical platforms,
including photonic
(both `discrete-variable'~\cite{BPM+97,BBD+98} and `continuous-variable'~\cite{FSB+98}),
nuclear magnetic resonance~\cite{NKL98},
optomechanical~\cite{HNA+21},
solid-state systems~\cite{PHB+14}, trapped atoms~\cite{NNR+13} and trapped ions~\cite{RBB+24}.
Teleportation of qudits has also been reported~\cite{LZE+21}.
Experimental performance is typically benchmarked according to the average gate fidelity~\cite{BFK00}.
Whether quantum teleportation has been achieved in any given experiment can be controversial~\cite{BK98,RS01a,RS01b,vEF01,Wis03,NB04,Smo04},
which is a key motivation for this work.
\subsection{State fidelity}
\label{subsec:fidelity}
Fidelity is an important figure of merit for assessing the performance of quantum operations.
We elucidate state fidelity and how to estimate average state fidelity,
then discuss how~$f_\text{\text{th}}$ is established.

Fidelity quantifies the proximity between two quantum states by computing the Hilbert-space inner product~\cite{NC10}.
For our purposes,
one of the states is always pure, hence~$\ket\psi$,
and the other state could be pure or mixed,
hence~$\rho$.
The fidelity between these pure and mixed states is
\begin{equation}
\label{eq:fpsi}
f:\mathcal{B}(\mathscr{H})\times\mathscr{H}\to[0,1]:
(\rho,\ket\psi)\mapsto\expval{\rho}{\psi},
\end{equation}
with 
\begin{equation}
    \label{eq:fpsiphi}
    f=\abs{\braket{\psi}{\phi}},
\end{equation}
for pure~$\rho=\ket\phi\bra\phi$.
The extremes
\begin{equation}
\label{eq:extreme}
f=1\Leftarrow\rho=\ket\psi\bra\psi,\,
f=0\Leftarrow\rho=\ket\psi^\top\!\bra\psi,
\end{equation}
for~$^\top$ denoting complement;
i.e.,
$\bra{\psi}\ket{\psi}^\top=0\forall\ket{\psi}^\top\in\mathscr{H}$.
For pure~$\rho=\ket\phi\bra\phi$,
\begin{equation}
\label{eq:unitaryphi}
\ket\phi=R(\phi)\ket0,\;
R(\phi)\in\mathcal{U}(\mathscr{H})\cong\operatorname{SU}(d),\,
d=\text{dim}(\mathscr{H})
\end{equation}
with~$R$ used to denote unitary gates as the term `rotation' is typically employed,
and~$\ket0$ is some agreed canonical state (not necessarily a logical zero state) in~$\mathscr{H}$. The transformation $R(\phi)$~(\ref{eq:unitaryphi}) has $d^2-1$ real parameters~\cite{Wyb74}.

For~$\{\psi\}$ an appropriate parameterisation of states,
the uniform measure is denoted~$\text{d}\mu(\psi)$,
and average state fidelity is
\begin{equation}
\label{eq:barf}
\Bar{f}(\rho)
:=\int\text{d}\mu(\psi)\expval{\rho}{\psi}\in[0,1],\,
\int\text{d}\mu(\psi)=1.
\end{equation}
For pure state 
\begin{equation}
\label{eq:purerho}
    \rho =\ket\phi\bra\phi,
\end{equation}
the average state fidelity is re-expressed as
\begin{equation}
\label{eq:barf1}
\Bar{f}(\rho)
:=\int\text{d}\mu(\psi)\abs{\braket{\psi}{\phi}}^2.
\end{equation}
Integrals reduce to sums in discrete cases.

Now we restrict to the single-qubit case.
Any single-qubit state~$\ket\psi$ 
is generated from canonical~$\ket0$
by applying
\begin{equation}
R(\psi)\in\mathcal{U}(\mathscr{H})\cong\operatorname{SU}(2).
\end{equation}
Typically, we fix the coefficient of logical~$\ket0$ to be real
(the projective part of~$\mathbf{CP}^2$)
so
\begin{equation}
\label{eq:SU2U1}
R(\theta,\phi)
\in\text{SU}(2)/\text{U}(1)\cong\mathcal{S}^2
\end{equation}
with~$\mathcal{S}^2$ the Bloch sphere
parameterised by polar angle~$\theta$ and azimuthal angle~$\phi$.
Average state fidelity is thus obtained by using the uniform measure
\begin{equation}
\label{eq:uniformmeasure}
\dd\mu(\psi)\mapsto-\dd(\cos\theta)\dd\phi/4\pi=:\text{d}\Omega
\end{equation}
on the sphere, leading to
$\Bar{f}:=-\oiint\text{d}\Omega f(\Omega)$~\cite{CB07}
with the closed double integral being over the entire (Bloch) sphere.

Average state fidelity is a standard tool for evaluating whether experimental quantum teleportation has been successful or not based on surpassing some $f_\text{\text{th}}$.
The argument for assessing success or otherwise by averaging state fidelity is based on regarding an unknown state as sampling from a known distribution of states and quantifying performance by average~\cite{BFK00}.
Our approach is to justify characterisations such as average gate fidelity in terms of an adversary model.
Entanglement fidelity is a measure used to assess the process of teleporting an entangled state~\cite{JBS02}.
Consider the superoperator $\mathcal{E}$ acts linearly on the input density matrix $\rho$, producing an output state $\mathcal{E}(\rho)$.
The entanglement fidelity 
\begin{equation}
    \label{eq:entf}
    f_{\text{e}}:=\bra{\Psi} (\mathds{1} \otimes \mathcal{E})(\ketbra{\Psi})\ket{\Psi}
\end{equation}
In the case of pure qubits, entanglement fidelity simplifies to the standard fidelity~(\ref{eq:fpsi}).

The fidelity of a quantum state is not directly measurable;
instead an approach such as quantum state tomography is used to estimate the state~\cite{Schmied16}.
Quantum state tomography is achieved by performing numerous measurements over a tomographically complete set of measurement operators~\cite{NC10}
and then aggregating the data to estimate the state fidelity.

\begin{comment}
\paragraph{\color{red} State tomography:}\textcolor{blue}{Here we introduce quantum state tomography. Quantum state tomography is the procedure of experimentally
determining an unknown quantum state~\cite{NC10}.
For a single qubit, multiple copies of the unknown state \begin{equation}
    \rho=\nicefrac12 (1+\Vec{r}.\Vec{\sigma}),
\end{equation} 
are prepared.
After, the expectation value of $\sigma_i$, where $i\in \{0,\dots,3\}$, is calculated to find the estimates of the block vector $\bar{r_i}$ 
}
\end{comment}

\subsection{Quantum-circuit diagram}
\label{subsec:qcircuitdiagram}
In this subsection, we explain a quantum circuit and its components including operations, unitary gates and gadgets.
This explanation is needed as we rely heavily on quantum circuits to evaluate~$f_\text{\text{th}}$ for quantum teleportation.
A quantum circuit representing a unitary transformation on qubits can be decomposed into an agreed set of standard unitary logical gates~\cite{DN05}.
In Fig.~\ref{fig:operations},
a quantum circuit is represented pictorially,
called a quantum-circuit diagram,
by assigning horizontal wires to qubits with time increasing to the right, 
and we follow the convention that unlabelled wires are initialised to~$\ket0$;
gates (boxes on wires),
measurements (shown as a meter)
and outputs (right of the wires) are shown.

Now we focus on unitary gates,
which are shown in Figs.~\ref{fig:operations} (a), (c), and (d).
We begin by describing specific single-qubit rotation gates used in our quantum circuits,
then generalise to multi-qubit rotation gates
and to unitary `gadgets', which are convenient concatenations of unitary gates.
As a special case we explain the CNOT gate.

Special cases of single-qubit rotation gates are depicted in Figs.~\ref{fig:operations}(a) and~(d)
with the gates described in the figure caption.
Note that the Hadamard gate~\cite{NC10}
\begin{equation}
\label{eq:Hadamard}
H\ket \varepsilon\mapsto(\ket0 +(-1)^\varepsilon\ket1)/\sqrt2,\,
\varepsilon\in\{0,1\}.
\end{equation}
The matrix representation for Hadamard gate is
\begin{equation}
H=\frac1{\sqrt{2}}\begin{pmatrix}
1 & 0 &\\
0& -1 &
\end{pmatrix},
\end{equation}
has the property that $H\notin\text{SU}(2)$
due to the determinant being $-1$ rather than $1$.
However, this problem is ignored as this negativity cancels out in circuits of interest.

Single-qubit Pauli gates (X, Z, Y) are other cases not shown in Fig.~\ref{fig:operations}.
We represent a multi-qubit state by
\begin{equation}
\label{eq:multiqubit}
   \ket\Psi_m \in\text{span} \{\ket{\bm\varepsilon}_m;
    \bm\varepsilon\in\{0,1\}^*\}
\end{equation}
with~$^*$ wildcard notation indicating any size Cartesian product of~$\{0,1\}$;
the state~(\ref{eq:multiqubit}) is a generalised single-qubit state~(\ref{eq:qubit}).
The $m$-qubit rotation gate is 
\begin{equation}
\label{eq:multiqubitrotation}
\text{SU}(2^m)\ni R(\Psi):
\ket0^{\otimes m}\mapsto\ket\Psi_m,
\end{equation}
which can be decomposed into SU(2) transformations~\cite{RZBB94,dGSBZ01}.
If~$m=1$, we write $\ket\Psi_{m=1}=\ket\psi$.
The CNOT gate is  
\begin{equation}
\label{eq:cnot}
\text{CNOT}\ket{\varepsilon_0\varepsilon_1}\mapsto\ket{\varepsilon_0~\varepsilon_1\oplus \varepsilon_0}.
 \end{equation}
 The CNOT gate matrix representation
\begin{equation}
\text{CNOT} =\begin{pmatrix}
1 & 0 & 0 & 0\\
0 & 1 & 0 & 0\\
0 & 0 & 0 & 1\\
0 & 0 & 1 &0
\end{pmatrix}
\end{equation}
CNOT is a quantum version of the logical XOR gate~\cite{rabaey2002digital}.

Unitary gates over multiple qubits can be composed to form a unitary gadget.
Entanglement gadgets are especially useful as a component both in preparing entangled states~\cite{NC10} and in performing joint entanglement measures such as a Bell-inequality test~\missingcite.
As an example,
the canonical entanglement gadget is mathematically~\cite{NC10}
\begin{align}
\label{eq:entgadget}
\text{CNOT}&(H\otimes\mathds 1)\ket{\varepsilon_0\varepsilon_1} \nonumber\\
=& \frac{\text{CNOT}}{\sqrt{2}} \left(\ket{0\varepsilon_1} +(-1)^{\varepsilon_0}\ket{1\varepsilon_1}\right )\nonumber\\ =&\left(\ket{0\varepsilon_1} +(-1)^{\varepsilon_0}\ket{1~\varepsilon_1\oplus 1}\right)/\sqrt2.
\end{align}
The matrix representation for the entanglement gadget is
\begin{equation}
\label{eq:matD}
    \mathscr{E} =\frac1{\sqrt{2}}\begin{pmatrix}
      1 & 0&1 &0 \\
      0&1 &0 &1   \\
     0 & 1& 0&-1 &  \\
        1& 0& -1&  0\\
    \end{pmatrix}
\end{equation}
This entangling gate is important as a universal quantum
computation can be achieved using only qubit rotations and two-qubit entangling gates~\cite{dodd2022}.
The entanglement-gadget inverse is
\begin{align}
\label{eq:inverseentgadget}
(H\otimes\mathds 1)&\text{CNOT} \ket{\varepsilon_0\varepsilon_1}\nonumber\\
=& (H\otimes\mathds 1)\ket{\varepsilon_0~\varepsilon_0\oplus \varepsilon_1}\nonumber\\ =& (\ket{0~\varepsilon_0\oplus \varepsilon_1}
+(-1)^{\varepsilon_0}\ket{1~\varepsilon_0\oplus \varepsilon_1\oplus 1})/\sqrt{2},
\end{align}
which is employed in a BSM.

Now that we have explained the unitary elements of a quantum circuit,
here we briefly describe the quantum circuit's non-unitary elements.
Mathematically,
these generalised transformations are operations,
which are completely positive, non-tracing increasing map.
We first explain qubit measurement, 
followed by trashing a qubit. Finally, we define a random-bit generator,
which plays a key role in spoofing quantum teleportation.

Here we define a destructive multi-qubit measurement,
with `destructive' meaning that the quantum state is removed and replaced by a number in constrast to a positive operator-valued measure that maps operators to operators~\missingcite.
We specify a destructive qubit measurement in a restricted form that suffices for our purposes:
Specifically, a multi-qubit measurement is a tensor product of single-qubit measurements.
Capturing the inherent randomness of measurement in quantum mechanics,
a single-qubit measurement is achieved in our circuit diagram by mapping a qubit intoa random Bernoulli variable, which is a random process labelled by some~$r\in(0,1)$.
We express this measurement process mathematically as
\begin{equation}
\label{eq:1qubitmeasurement}
M:
\mathscr{H}_2\to\{0,1\}:
\ket\psi\mapsto\begin{cases}
0 & \text{if unif}[0,1]\geq r,\\
1 & \text{otherwise},   
\end{cases}
\end{equation}
with unif referring toa uniform distribution
and $[0,1]\subset\mathbb{R}$
is the unit interval of the real numbers.
Note that unif in Eq.~(\ref{eq:1qubitmeasurement}) implies each outcome is uniformly random according to instances of random distribution.
This random process,
with
\begin{equation}
\label{eq:r}
r=|\psi_0|^2,
\end{equation}
is depicted in Fig.~\ref{fig:operations}(b)
for the qubit~(\ref{eq:qubit}) 

Now we explain how to `trash' a qubit.
Our QTPs involve trash operations, which discard qubits and collect no information.
We represent a discarded qubit by~$\emptyset$,
meaning that the qubit is not just erased but no longer exists in the system.
Therefore, trash is different from measurement where the qubit can be discarded, but a classical measurement outcome is obtained.
Trash is mathematically a partial trace
\begin{equation}
\label{eq:partialtrace}
\tr_m:\mathscr{H}^m\to\mathscr{H}^{m-1},
\end{equation}
which removes the $m^{\text{th}}$ Hilbert space in the tensor product, as shown in Fig.~\ref{fig:operations}(b).

Now we introduce a random-bit generator.
Random-bit generators are employed in the classical literature, 
but an explicit incorporation of random-bit generation into quantum circuits is novel here.
In Fig.~\ref{fig:operations}(e),
we show a classical random-bit generator~$G$ with a double-line input (single bit) and a double-line output. 
This~$G$ is important in our adversarial setting involving cheating.

For~$G$,
the input is a seed,
which is some bit string
that is transformed according to a mathematical formula~\cite{BBS86}.
This mathematical formula has the effect of rapidly converting a single input seed value into an effectively uniform Bernoulli variable~(\ref{eq:1qubitmeasurement})
\begin{equation}
\label{eq:RBG}
G:0\mapsto \begin{cases}
0 & \text{if unif}[0,1]\geq \nicefrac12,\\
1 & \text{otherwise}.  
\end{cases}
\end{equation}
with $r=\nicefrac12$.
\section{Approach}
\label{sec:approach}

In this section, we describe our approach to establishing~$f_\text{\text{th}}$ for various QTPs, with different QTPs corresponding to different cases involving honest and cheating agents. First, in~\S\ref{subsec:model}, we present our model,
which allows for any one of four QTPs involving four agents in each QTP.
Three of these four QTPs
allow two agents,
A and/or~B
to cheat,
and the other protocol is for honest~A and~B.
We then discuss certification to guarantee that the QTP has been implemented honestly,
with this success depending on achieving or surpassing a particular~$f_\text{\text{th}}$.
Then
in~\S\ref{subsec:math},
we describe our mathematical analysis of the agents' role in the honest protocol and cheating protocols. 
\subsection{Model}
\label{subsec:model}

In this subsection, we introduce our model for each of the five protocols that involve honest and/or cheating agents. 
First, we define our protocols, starting with defining the role of each agent, followed by describing the honest protocol, and then we describe the cheating protocols.
Finally, we discuss the certification procedure, which depends on evaluating~$f_\text{\text{th}}$.

We begin by describing our four-agent QTPs.
Initially, we describe all agents and their roles, regardless of whether the agent is honest or not.
Subsequently, we describe the various QTPs with all agents honest, and then either~A or~B cheats, or both~A and~B cheat together.
Note that we use the term `protocol' to refer to the rules not only for the honest parties but also for the cheating parties, so each cheating strategy can have its own QTP.
Although creating a protocol for cheating is initially surprising,
our cheating QTPs are used to set benchmark standards, so we need clearly defined QTPs even for cheating protocols.

Our QTP employs four agents, namely, the certifier~C, the sender~A, the receiver~B, and the entanglement supplier~D.
D's~role is to prepare an entangled state, send one share to~A and the other share to~B.
A's~task is to send quantum information through a classical channel to~B, which is achieved by a performing two binary measurements and announcing publicly 
\begin{equation}
\label{eq:2bits}
(a,b)\in\{0,1\}^2
\end{equation}
which~B receives.
B's~role is to use the received classical information to reconstruct the original state.

C~receives from~B the state that should ideally be a reconstruction of the original state that was created by~C. 
In our approach, C~does not hear the announcement $(a,b)$. We treat QTP as a black box in which the announcement is restricted to A~and~B.
If, instead, $(a,b)$ was announced, our threshold condition would be different.

Now we explain how~C issues the certificate based on estimating the fidelity of the reconstructed state with respect to the original state, without knowing the announcement $(a,b)$.
C's task is to certify that QTP is successful despite~A and/or B being adversarial,
meaning that~A and/or B cheat by spoofing quantum teleportation using classical resources.
Thus, C's certificate,
issued only if C's estimate of the efficacy of the protocol, according to an appropriate measure, exceeds a predetermined threshold value based on the adversarial model.
If the threshold condition is met or surpassed,
C might issue a certificate as follows.
\begin{cert}\label{cert:p0}
  This system executes the QTP perfectly.  
\end{cert}
\noindent
This certificate~\ref{cert:p0} guarantees that
A \&\ B have undertaken their respective tasks both perfectly and honestly.
Of course such a certificate would never be issued as perfection is unattainable,
but this certificate illuminates how to write alternative certificates for the cheating protocols.

Our QTP relies on representing quantum teleportation as a quantum-circuit diagram and modifying this quantum circuit according to our cheating protocols.
Now we describe the honest QTP denoted by $\wp^\text{0}$
(with~$\wp$ shorthand for QTP
and~`0' referring to the $0^\text{\text{th}}$, or honest, QTP)
depicted in Fig.~\ref{fig:pr0}.
$\wp^\text{0}$ is pertinent to the original QTP,
except that the original QTP actually only specifies agents~A and~B and treats entanglement as input rather than provided by an agent~D as in our case.

In our case, we explicitly describe the additional two agents~C and~D.
For~$\wp^\text{0}$,
C creates an $m$-qubit state ($m$~qubits)
\begin{equation}
\label{eq:ketPsi}
\ket{\Psi}_m:=R(\Psi)\ket0^m,
\end{equation}
which is expressed the same for all protocols, not only for $\wp^\text{0}$.
We then provide~A with a share
as represented by a quantum circuit diagram shown in Fig.~\ref{fig:pr0}. 
\begin{figure}
\begin{tcolorbox}
 \[
\begin{array}{c}
\Qcircuit @C=1.2em @R=1.5em {
 &&&\mbox{C} & &  &&&\\
 &{/^{m-1}}\qw&\qw&\multigate{1}{R(\Psi)}  & \qw&\qw &{/^{m-1}}\qw& \qw& \qw&\qw&\qw\\
&\qw&\qw& \ghost{R(\Psi)}&\qw&\ctrl{1} &\gate{H}&\qw&\meter&\mbox{A}&&\mbox{C}\\
 &\qw&\qw& \gate{H} & \ctrl{1} &\targ& \qw&\meter&&&\\  &\qw&\qw& \qw & \targ & \qw&\qw&\gate{X}\cwx &\gate{Z}\cwx[-2] &\qw&\qw\\
& &&\mbox{D} &~&~~~ &&~~~\mbox{B}\gategroup{3}{6}{4}{9}{1.1em}{--}\gategroup{5}{8}{5}{9}{.75em}{--}\gategroup{3}{4}{2}{4}{.8em}{--}
\gategroup{4}{4}{5}{5}{.75em}{--}
\gategroup{2}{11}{5}{11}{.75em}{--}}
\end{array}\]
\end{tcolorbox}
\caption{%
Quantum circuits representing teleporting a one-qubit share of an $m$-qubit state ($\wp^\text{0}$).
The input is~$\ket0^{m+2}$; the absence of a label on the left side implies that each input is~$\ket0$.
The three lower horizontal lines (`wires') each represent one qubit, and the uppermost horizontal line with a slash drawn through it corresponds to $m-1$ qubits, which is a label on the slash.
This slashed wire is for ancillary qubits.
C~performs the $R(\Psi)$ operation, which is an $m$-qubit gate parameterised by $\Psi$ (a string of parameters).
D~effects the entangling gadget~(\ref{eq:entgadget}), providing a maximally entangled pair of qubits (ebit).
Agents A, B, C and~D are depicted by dashed boxes illustrating the operations they execute.
At the end~B sends the qubit to~C, which does not have a dashed box. Finally, C~holds the qubit sent from~B and holds the ancillary qubits as well.
}
\label{fig:pr0}
\end{figure}
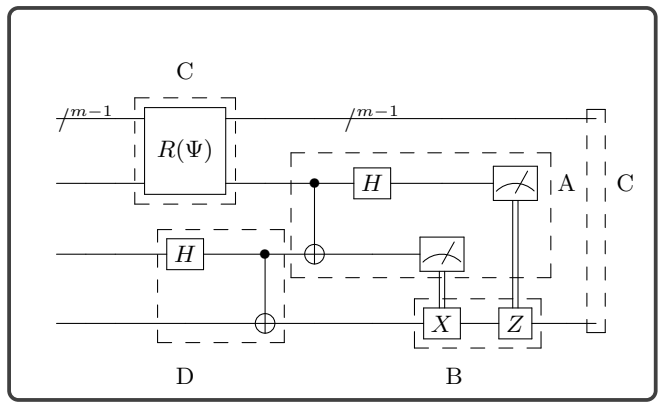
A simpler case of~$\wp^\text{0}$ corresponds to~C sending the entire state to~A.
Then, for~$\wp^\text{0}$, D sends ~$\ket{\widetilde{\bm 0}}$
(\ref{eq:ebit}) to both~A and~B.
Both~A and~B behave as described in \S\ref{subsec:qteleportation}. 
Finally, B~sends his qubit to~C, who holds the ancillary qubits.

Now we describe our QTPs involving cheaters (cheating agents).
These QTPs comprise two QTPs for cheating~A and honest~B
(with `honest' meaning that~B performs the task as instructed but not necessarily perfectly),
one QTP for  honest~A and cheating~B, and one QTP for A~and~B both cheating. 
In the first two cheating QTPs, A~is the sole cheater and follows either of the two QTPs~$\wp^\text{A1}$ or~$\wp^\text{A2}$.
These two cases are represented by quantum circuit diagrams in Fig.~\ref{fig:prA}.
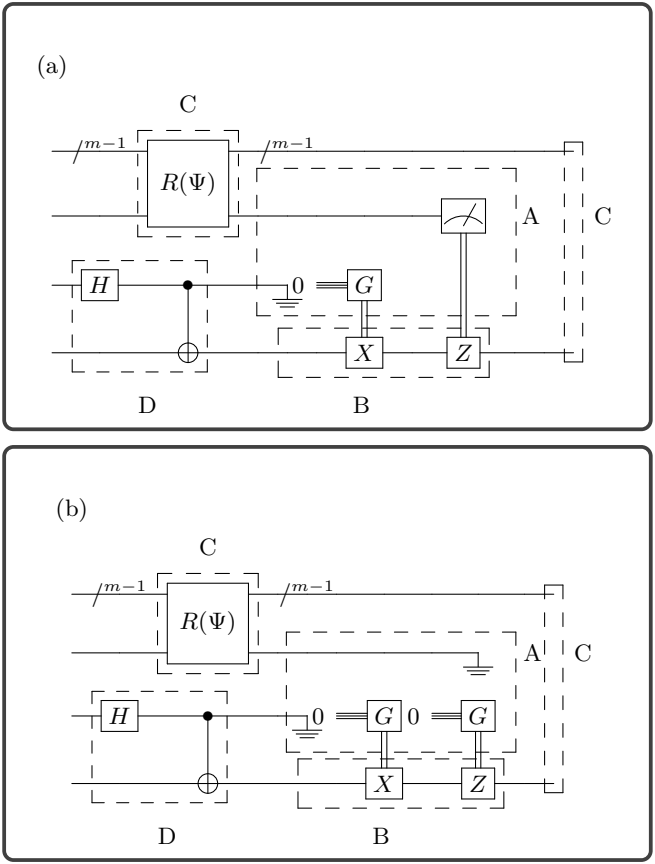
\begin{figure}
\begin{tcolorbox}
\[
\begin{array}{c}
\Qcircuit @C=1.2em @R=1.5em {
 \mbox{(a)} &&  &&&\\
 & & \mbox{C} &  &&&\\
 &{/^{m-1}}\qw& \multigate{1}{R(\Psi)} &\qw & {/^{m-1}} \qw&\qw&\qw& \qw&\qw&\qw&\qw&\qw\\
&\qw& \ghost{R(\Psi)}&\qw&\qw&\qw&\qw&\qw&\meter&~~~~\mbox{A}&&&\mbox{C}\\
 & \gate{H} & \ctrl{1} & \qw&\ground \qw& \lstick{0}&\gate{G}\cw&&&&\\  & \qw & \targ & \qw&\qw&\qw&\gate{X}\cwx & \qw&\gate{Z}\cwx[-2] &\qw&\qw&\qw\\
& &\mbox{D} ~~~~~~~~~~&~~~ &~~~&~~~~~~~~~~~\mbox{B}
\gategroup{4}{5}{5}{9}{2.5em}{--}
\gategroup{6}{5}{6}{9}{.75em}{--}
\gategroup{3}{3}{4}{3}{.8em}{--}
\gategroup{5}{2}{6}{3}{.75em}{--}
\gategroup{3}{12}{6}{12}{.75em}{--}}
\end{array}\]
\end{tcolorbox}
\begin{tcolorbox}
 \[
\begin{array}{c}
\Qcircuit @C=1.2em @R=1.5em {
\mbox{(b)} &&&&&\\
 & & \mbox{C} & & & &\\
 &{/^{m-1}}\qw& \multigate{1}{R(\Psi)} & \qw &{/^{m-1}}\qw&\qw&\qw& \qw&\qw&\qw&\qw\\
&\qw& \ghost{R(\Psi)}&\qw&\qw&\qw&\qw&\qw&\ground \qw&~~\mbox{A}&&\mbox{C}\\
 & \gate{H} & \ctrl{1} & \qw&\ground \qw& \lstick{0}&\gate{G}\cw& \lstick{0}&\gate{G}\cw&&\\  & \qw & \targ & \qw&\qw&\qw&\gate{X}\cwx & \qw&\gate{Z}\cwx[-1] &\qw&\qw\\
& &\mbox{D} ~~~~~~~~~~&~~~ &~~~&~~~~~~~~~~~\mbox{B}
\gategroup{4}{5}{5}{9}{1.7em}{--}
\gategroup{6}{5}{6}{9}{.75em}{--}
\gategroup{3}{3}{4}{3}{.8em}{--}
\gategroup{5}{2}{6}{3}{.75em}{--}
\gategroup{3}{11}{6}{11}{.75em}{--}}
\end{array}\]
\end{tcolorbox}
\caption{%
Quantum circuits represent two A~cheating protocols:
(a)~$\wp^\text{A1}$ wherein~A trashes out her second qubit and replaces her operation with a random bit generator $G$. 
(b)~$\wp^\text{A2}$ with~A trashing both qubits and replaces both qubits with random bit generators $G$. 
In both QTPs, A~sends the two qubits to~B. 
Here C, D, and B follow the same procedure as in Fig.~\ref{fig:pr0},
and the constructs such as the wires and slashed wire are the same as for that figure as well.}
\label{fig:prA}
\end{figure}

For~$\wp^\text{A1}$, A~ measures her first qubit and trashes her second qubit shown in Fig.~\ref{fig:prA}(a).
Whereas, for~$\wp^\text{A2}$, A~trashes both of her qubits shown in Fig.~\ref{fig:prA}(b). 
When~A trashes a qubit, she replaces it with a random bit.
Then, A~sends the two bits to~B. 

For the next QTP, namely, honest~A and cheating~B,
denoted~$\wp^\text{B}$, B~trashes one of (honest) A's second bits and~D's qubit. 
Finally,~B applies X to~$\ket 0$ only if A's first qubit is measured as~1. This protocol is represented by a quantum circuit diagram shown in Fig.~\ref{fig:prB}.
\begin{figure}
\begin{tcolorbox}
 \[
\begin{array}{c}
\Qcircuit @C=1.05em @R=1.5em {
 &&& \mbox{C}&  & & & \\
 &{/^{m-1}}\qw&\qw& \multigate{1}{R(\Psi)} &\qw & {/^{m-1}} \qw&\qw&\qw& \qw&\qw& \qw\\
&\qw&\qw& \ghost{R(\psi)}& \qw  & \ctrl{1} &\gate{H}&\qw & \meter &\mbox{A}&&\mbox{C}\\ &\qw&\qw&\gate{H} & \ctrl{1}& \targ & \qw&\meter&&\\  &\qw&\qw& \qw & \targ & \ground \qw&  &\gate{\mathds1}\cwx[-1] &\gate{X}\cwx[-2] & \qw& \qw\\
& &&&\mbox{D} ~~~~~~~~~~&~~~ &~~~&~~~~~~~~~~~\mbox{B}
\gategroup{3}{6}{4}{9}{.8em}{--}
\gategroup{5}{6}{5}{9}{1.5em}{--}
\gategroup{2}{4}{3}{4}{.8em}{--}
\gategroup{4}{4}{5}{5}{.75em}{--}
\gategroup{2}{11}{5}{11}{.75em}{--}} 
\end{array}\]
\end{tcolorbox}
\caption{%
Quantum circuit representing $\wp^\text{B}$. B~trashes his qubit and replacew his X-gate by~$\mathds{1}$ and his Z-gate with by an X-gate.}
\label{fig:prB}
\end{figure}
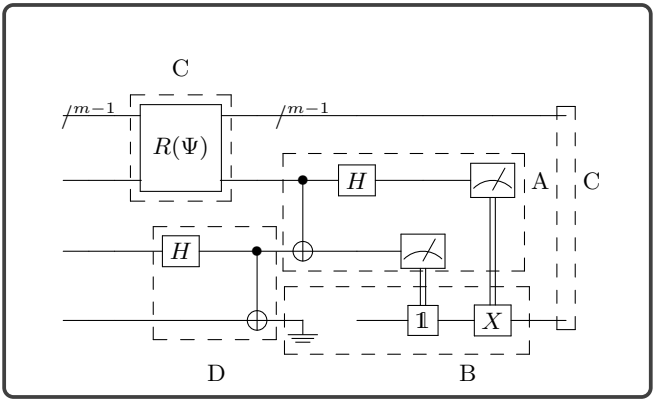
For the QTP with both~A and~B cheating,
denoted~$\wp^\text{AB}$ and represented by a quantum circuit diagram shown in Fig.~\ref{fig:prAB},
\begin{figure}
\begin{tcolorbox}
 \[
\begin{array}{c}
\Qcircuit @C=1.4em @R=1.5em {
 & &&&&\mbox{C}&  & & & &\\
 &{/^{m-1}}\qw&\qw& \multigate{1}{R(\Psi)} & \qw & {/^{m-1}}\qw& \qw&\qw&\qw & \qw& \qw& \qw\\
&\qw&\qw& \ghost{R(\psi)}& \qw & \qw& \qw&\qw&\qw&\meter& ~~\mbox{A}&&\mbox{C} \\ &\qw&\qw& \gate{H} & \ctrl{1} & \qw& \ground \qw&&&\\  &\qw& \qw& \qw & \targ & \qw& \ground \qw&  & &\gate{X}\cwx[-2] & \qw& \qw\\
& &&&\mbox{D} ~~~~~~~~~~&~~~ &~~~&~~~~~~~~\mbox{B}
\gategroup{2}{4}{3}{4}{.8em}{--}
\gategroup{3}{7}{4}{10}{2.3em}{--}
\gategroup{4}{4}{5}{5}{.75em}{--} \gategroup{5}{7}{5}{10}{1.5em}{--}
\gategroup{2}{12}{5}{12}{.75em}{--}} 
\end{array}\] 
\end{tcolorbox}
\caption{%
Quantum circuit representing $\wp^\text{AB}$.
A~and~B trash their qubits.}
\label{fig:prAB}
\end{figure}
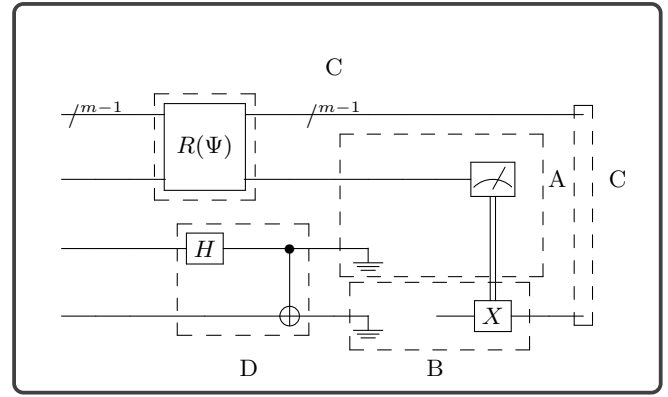
A~measures the qubit from~C and then trashes the qubit from~D. 
Then B~applies X to~$\ket 0$ only if A's first qubit is measured as~1.

C evaluates the performance of the protocol to decide whether to issue the certificate.
The typical way to determine how well quantum teleportation,
or other quantum tasks such as entanglement swapping~\cite{ZZHE93} and quantum logic gates for quantum computing~\cite{NC10},
is achieved
involves assessing the fidelity of the implementation relative to the target input state~(\ref{eq:fpsi}).

Now we explain the threshold fidelity.
For our QTPs~(\ref{eq:QTPs}),
A announces a two-bit measurement result,
and C then assesses the quantum state fidelity for each of the four cases.
C issues the certificate if a threshold-fidelity condition is met,
with the threshold fidelity~$f_\text{th}$
a sum of fidelities for each of the four announcements
\begin{equation}
\label{eq:announcements}
(a,b)\in\{0,1\}^2
\end{equation}
Our approach to determine this weighted sum for each QTP~(\ref{eq:QTPs});
if~$f_\text{th}$ exceeds this weighted sum for a given adversarial QTP,
then~C issues a certificate saying that this QTP necessarily involved~A and~B using quantum resources.
Note that using the~$f_\text{th}$ condition does not guarantee that cheaters are selectively deleting some instances based on~$(a,b)$
announcements.

Now consider~$\wp^{\text{A}}$
with~A cheating and~B honest.
If all parties execute perfectly,
the analogue of Certificate~\ref{cert:p0} is the following,
which replaces Certificate~\ref{cert:perfectpA}.
\begin{cert}
[Perfect $\wp^{\text{A}1}$ and $\wp^{\text{A}2}$]
\label{cert:perfectpA}
A~employs more than zero qubits from~D for the QTPs.
\end{cert}
\noindent
Of course such a certificate is useless, both because it can only be issued if perfection is attained and also because a certificate that someone is cheating is not useful.
Instead, 
the certificate should be that~A is \emph{not} cheating.
The useful certificate is then the following.
\begin{cert}[Practical $\wp^{\text{A}1}$ and $\wp^{\text{A}2}$]
\label{cert:pA}
This system executes the QTP with~A having used quantum resources and~B assumed to have used quantum resources.
\end{cert}
\noindent
For this certificate,
B is trusted to be using quantum resources to perform his share of the protocol honestly.
The certificate does not ascertain that~A has performed her task honestly but rather that she has expended quantum resources for the threshold-fidelity condition to have been achieved.
This certificate is essentially the one employed for experimental continuous-variable quantum teleportation~\cite{braunsteinK98}
and its justification as successful quantum teleportation~\cite{BFK01},
although the continuous-variable context vs our discrete-variable context brings in subtleties beyond the scope of our work.

We finish by introducing two more certificates,
one for honest~A and cheating~B ($\wp^{\text{B}}$)
and the other for
both~A and~B cheating~($\wp^{\text{AB}}$).
The first certificate is the following.
\begin{cert}[$\wp^{\text{B}}$]
\label{cert:pB}
This system executes the QTP with~A assumed to have used quantum resources and~B having used quantum resources. 
\end{cert}
\noindent
This certificate does not assure that~B has performed his task honestly.
Here,~A is trusted to use quantum resources to perform her share of the QTP honestly.
\begin{cert}[$\wp^{\text{AB}}$]
\label{cert:pAB}
This system executes the QTP with both~A and~B having used quantum resources. 
\end{cert}
\noindent
Now we proceed to present our mathematical calculations.
\subsection{Mathematics}
\label{subsec:math}
In this subsection, we present our mathematical description of the model~\ref{subsec:model}.
We begin by describing our four agents mathematically, specifically by introducing linear operators whose domain includes both Hilbert space  and probability space, 
for QTPs
\begin{equation}
\label{eq:QTPs}
\wp^\text{0},\,
\wp^\text{A1},\,
\wp^\text{A2},\,
\wp^\text{B},\,
\wp^\text{AB}.
\end{equation}
Then we describe the output states for different cheating QTPs. 

Now we proceed to the mathematical description of the four agents for the honest QTP~$\wp^\text{0}$.
Our description employs pure states and unitary maps so is for an ideal case.
As we deal only with pure states here,
we avoid partial trace,
which yields mixed states,
and instead write the action as a unitary map on all qubits, not just the agent's.
This unitary map explicitly includes writing the identity operator~$\mathds1$
acting that acts on all qubits outside the domain of the respective agent.
The notation~$\mathds1^m$
indicates the identity action on~$m$ qubits.

We begin with a mathematical description of~C's role in the QTP of Fig.~\ref{fig:pr0}.
C applies the general
$m$-qubit rotation gate
\begin{equation}
\label{eq:C} 
R(\Psi)\otimes\mathds1^2
\in\mathcal{B}\left(\mathscr{H}_2^{m+2}\right),\,
\mathds1^k:=\underbrace{\mathds1\otimes\mathds1\otimes\cdots\otimes\mathds1}_{k\text{ times}},
\end{equation}
to~$\ket0^m$,
which generalises $R(\Psi)$~(\ref{eq:multiqubitrotation}) by appending an identity transformation on ancillary qubits.
Note that 
\begin{equation}
    \label{eq:trid}
    \tr\mathds1^k =\tr \left( \sum_{i\in \{0,1\}^k}\ket i\bra i\right) = 2^k
\end{equation}
with
\begin{equation}
\label{eq:trpure}
\tr\left( \ket i\bra{i} \right)=\braket{i}{i}=1\;\forall i.
\end{equation}
In Eq.~(\ref{eq:C}),
~$\Psi$ 
is the $(m^2-1)$-dimensional parameter vector that labels an element of $\text{SU}(2^m)$,
which is the Lie group of isometries acting on the $(m+1)$-dimensional Hilbert space~$\mathscr{H}_2^{m+2}$.

For $m=1$,
C applies $R(\theta,\phi)$
\begin{equation}
\label{eq:bloch}
R(\theta,\phi)\ket0 =\cos\nicefrac{\theta}2\ket0+\text{e}^{\text{i} \phi} \sin\nicefrac{\theta}2\ket1,
\end{equation}
where
\begin{equation}
\label{eq:thetaphi}
\theta\in [0,2\pi],\,
\phi\in [0,2\pi].
\end{equation}
For~$m$ qubits,
C cannot apply every arbitrary rotation gate $R(\Psi)$ as the parameters belong to an uncountable set that requires perfect precision. Instead, C tests by selecting a subset of $R(\Psi)$.
 
Accordingly, we choose to restrict the analysis to certificates based on a one-parameter family of $m$-qubit Greenberger-Horne-Zeilinger states~\cite{GHZ89,Mer90}.
The rotation operation performed by~C on the initial state is
\begin{equation}
\label{eq:entop}
R(\Psi) =\cos\nicefrac\theta2 
\mathds1^{ m}+\sin\nicefrac\theta2 X^{m}
\end{equation}
for
\begin{equation}
\label{eq:Xm:=}
X^m:=X^{\otimes m}
\end{equation}
and the same for~$\mathds1^m$,
which includes the trivial case $\theta=0$, resulting in an input product state unchanged.
Generally, the mapping~(\ref{eq:entop}) interpolates from a~$\mathds1$ gate to a fan-out gate~\cite{Molmer_1999,Hoyer_2005} and then to a multi-qubit X-gate depending on~$\theta$. 

Next we describe mathematically D's action shown in Fig.~\ref{fig:pr0}.
As with~C,
we write the isometry over $m+2$ qubits so,
in this case,
D's action involves an $m$-qubit identity operator.
D executes the entangling gadget~(\ref{eq:entgadget})
as part of D's overall $(m+2)$-qubit transformation
\begin{equation}
\label{eq:D}
\mathds1^{ m}\otimes\left[ \left(\text{CNOT}\right)\left(H\otimes\mathds1\right)\right]
\in\mathcal{B}\left(\mathscr{H}_2^{m+2}\right),
\end{equation}
which,
as an example,
converts products states in the computational basis to
\begin{equation}
\label{eq:comutationalbasis}
\ket{\epsilon_0\epsilon_1},\,
\epsilon_\imath\in\{0,1\}
\end{equation}
into ebits $\ket{\widetilde{\epsilon_0\epsilon_1}}$~(\ref{eq:ebit}),
respectively.
D's operation on the state~(\ref{eq:entop})
yields
\begin{equation}
 \left(\cos\nicefrac\theta2 
\ket{0}^m+\sin\nicefrac\theta2\ket1^m\right)\otimes\ket{\widetilde{00}}
\in\mathscr{H}_2^{m+2},
\end{equation}
whose norm is unity.

Here we provide the mathematical description for honest A's action shown in Figs.~\ref{fig:pr0} and~\ref{fig:prB}. A~applies CNOT to her qubits with the $m^\text{\text{th}}$~qubit 
as the control (qubit sent from~C) and  the $(m+1)^{\text{\text{th}}}$~qubit as the target (qubit sent from~D) and~$\mathds1$ for the remaining qubits. Then A~applies H to the $m^\text{\text{th}}$~qubit. Finally, A~measures the~$m^\text{\text{th}}$ and~$(m+1)^{\text{\text{th}}}$ qubits, described by
\begin{equation}
\label{eq:A}
\mathds1^{m-1}\otimes\left(M_m\otimes M_{m+1}\right) \left(H\otimes\mathds1\right)\left(\text{CNOT}\right)\otimes\mathds1
\end{equation}
for~$M_k$
being the destructive measurement~(\ref{eq:1qubitmeasurement})
on the $k^\text{th}$ qubit
and the mapping being from $m+2$ qubits to~$m$ qubits plus two bits.
Then~A reports the pair of bits $(a,b)$~(\ref{eq:2bits})
to~B via classical two-bit communication.

Here we provide the mathematical description for cheating A's action for $\wp^{\text{A}1}$ and $\wp^{\text{A}2}$  shown in Figs.~\ref{fig:prA}(a,b), respectively. 
For $\wp^\text{A1}$, A~trashes the $(m+1)^{\text{\text{th}}}$~qubit (qubit sent from~D).
Then~A generates a bit~$0$ and applies G~(\ref{eq:RBG}).
Finally, A~measures the~$m^\text{\text{th}}$ qubit and reports the outcome of the measurement and the random bit number to~B. A's action is the replacement of the $(m+1)^\text{th}$ by a random bit.
Mathematically,
the destructive qubit measurement~$M_{m+1}$ in Eq.~(\ref{eq:A}) by
\begin{equation}
\label{eq:A1}
G_{m+1}(0)\circ \text{tr}_{m+1}
\end{equation}
where partial trace~(\ref{eq:partialtrace})
is the mathematical expression for trashing a qubit.
For $\wp^\text{A2}$, A~trashes the $m^\text{\text{th}}$ (qubit sent from~C) and the $(m+1)^{\text{\text{th}}}$~qubit (qubit sent from~D).
Then~A generates two $0$~bits and applies G and reports the two random bits to~B.
A's action on the two trashed qubits is thus
\begin{equation}
\label{eq:A2}
\left(G_{m}(0)\circ \text{tr}_{m}\right)\otimes \left(G_{m+1}(0)\circ \text{tr}_{m+1}\right)
\end{equation}
with random-number generation included.
 
For $\wp^\text{AB}$, A~trashes the $(m+1)^{\text{\text{th}}}$~qubit (qubits sent from~D).
Then~A measures the~$m^\text{\text{th}}$ qubit and reports the measurement output to~B.
A's action on her qubit,
which she destructively measures,
is
\begin{equation}
\label{eq:AB}
M_m\otimes \text{tr}_{m+1}
\end{equation}
which completes her task.
Honest B's action for $\wp^0$, $\wp^{\text{A}1}$ and $\wp^{\text{A}2}$ is
\begin{equation}
\label{eq:B}
\mathds1^{m-1}\otimes Z^{a}X^{b} \in\mathcal{B}\left(\mathscr{H}_2^{m}\right).
\end{equation}
depends on A's two bits $(a,b)$~(\ref{eq:2bits}).
For given~$(a,b)$,
the operator~(\ref{eq:B})
is unitary,
but cheating~A generates~$b$ randomly, and, for~$\wp^{\text{A}2}$ $a$ randomly,
so the resultant operator is essentially mixing unitary operators according to the probabilities of each choice of~$(a,b)$, namely~$p_ap_b$,
with this form holding because~$a$ and~$b$ are generated independently in all our scenarios,
including the cases where~$a$ or~$b$ results from destructive qubit measurements~(\ref{eq:1qubitmeasurement})
or by random-number generation~(\ref{eq:RBG}).
Consequently,
B's output state is mixed according to
\begin{equation}
\label{eq:mixedunitaryB}
\sum_{a,b\in\{0,1\}^2}
\underbrace{p_ap_bZ^aX^b\bullet\left(Z^aX^b\right)^\dagger}_{\rho_{ab}},
\end{equation}
with~$\bullet$
being the input to~B including all the qubits that~B does not deal with,
namely the qubits going to~A and~C as well,
hence~$m$ qubits. If $a$ and $b$ are equally likely to happen then 
\begin{equation}
    p_a=\nicefrac{1}{2},~~p_b=\nicefrac{1}{2} \implies p_ap_b=\nicefrac{1}{4},
\end{equation}
which implies that 
\begin{equation}
\label{eq:tracerhoab}
    \tr \rho_{ab}=\nicefrac14,
\end{equation}
if~$\bullet$ has a unit trace.
\begin{comment}
Here we describe B's actions for  $\wp^{\text{A}1}$ shown in Fig.~\ref{fig:prA}(a).
Bob's four resultant states,
for $m=1$, are
\begin{align}
\label{eq:psiabA1m=1}
\ket{\Psi_{00}}^{\text{A}1}
:=& \frac12\cos^2{\nicefrac{\theta}2}\big(\ket0\bra{0} \nonumber\\ 
&+\ket1 \bra1) \big)
\end{align}
\begin{align}
\label{eq:psiabA1}
\ket{\Psi_{00}}^{\text{A}1}_m:=& \frac12\cos^2{\nicefrac{\theta}{2}}\big(\ket0^{m} ~^{m}\bra{0} \nonumber\\ 
&+
 \ket0^{m-1} \ket1 (~^{m-1}\bra0 \bra1) \big)
\end{align}
\end{comment}

Next we describe cheating B's action for~$\wp^\text{B}$ QTP shown in Fig.~\ref{fig:prB}.
Here B~applies~$\mathds1$ to the $m^\text{\text{th}}$ qubit regardless of A's second measurement, which takes place as A's first qubit. Then~B applies either~$\mathds1$ and~$X$ gates to the $m^\text{\text{th}}$ qubit depending on whether~A reports 0 or 1 from her measurement of her first qubit. Mathematically, B's action is described by the operator,
\begin{equation}
\label{eq:B1}
 \mathds1^{m-1}\otimes  X^{a}\mathds1^{b} \in\mathcal{B}\left(\mathscr{H}_2^{m}\right),
\end{equation}
which completes his task.

Finally, shown in Fig.~\ref{fig:prAB} for the case that~A and~B collaborate,~B receives one bit from A's measurement of her first qubit and applies~$\mathds1$ or X gate to the $m^\text{\text{th}}$ qubit depending on whether that bit is 0 or 1 respectively. Mathematically, B's action
is described mathematically by
\begin{equation}
\label{eq:B2}
 \mathds1^{m-1}\otimes  X^{a} \in\mathcal{B}\left(\mathscr{H}_2^{m}\right).
\end{equation}
Now we proceed to describe the output states from the quantum circuits. 

We now describe the process for determining the output state for each of our cheating QTPs given the input~$\ket0^{m+2}$.
Mathematically,
C applies his operator~(\ref{eq:C}) to the first~$m$ qubits of the input state,
resulting in an $m$-qubit complex normalised vector.
At the same time, D~applies her operator~(\ref{eq:D}) to the last two qubits, resulting in a $2$-qubit vector.
Then~C sends his $m^\text{\text{th}}$ qubit to~A,
and D~sends her one of her shares as a qubit to~A and the other share as a qubit to~B.
Subsequently, A~applies her operator~(\ref{eq:A}) for~$\wp^\text{0}$ and~$\wp^\text{B}$ to her share of qubits received from~C and~D. 

In the cases of $\wp^\text{A1}$ and $\wp^\text{A2}$, A~applies the operators~(\ref{eq:A1}) and~(\ref{eq:A2}), respectively. For $\wp^\text{AB}$, A~applies a different operator~(\ref{eq:AB}).
As a result, the result of A's actions yields either two bits of information for $\wp^0$, $\wp^\text{A}$, and $\wp^\text{B}$, or just one bit for $\wp^\text{AB}$.
Based on A's output, B~applies the operator~(\ref{eq:B}) for $\wp^\text{0}$, a distinct operator~(\ref{eq:B1}) for $\wp^\text{B}$, and an alternative operator~(\ref{eq:B2}) for $\wp^\text{AB}$ to the qubit he received from~D.
Following the application of these operators, B~ends up with a single-qubit state. The output state of each quantum circuit is the qubit held by~B, which becomes entangled with the ($m-1$) qubit ancilla (the remaining $m-1$ qubits held by C).

Now we provide the mathematical description for the fidelity threshold required to certify that quantum resources were used in the execution of the QTP.
B delivers to~C his single-qubit share of the output state
averaged over announcements~$(a,b)$
for each of the honest QTP~$\wp^0$
and the four cheating QTPs~$\wp^{\text A1,A2},~\wp^{\text B},~\text{and}~\wp^{\text {AB}}$.
The fidelity threshold is
\begin{equation}
\label{eq:thresholdfidelity}
f_{\text{th}}:=\sum_{a,b\in\{0,1\}^2} \bra{\Psi}\rho_{ab}\ket{\Psi}
\end{equation}
for all QTPs.

\section{Results} 
\label{sec:results}
We present our results on~$f_\text{\text{th}}$
for each of the QTPs~(\ref{eq:QTPs}).
We first discuss how to establish~$f_\text{\text{th}}$ for $\wp^\text{0}$ in \S\ref{subsec:honestqtp}.
Then we present~$f_\text{\text{th}}$ for $\wp^{\text{A}1}$ and $\wp^{\text{A}2}$ in the case where A~cheats in \S\ref{subsec:acheatbhonest} and the case where~B cheats~$\wp^\text{B}$ in \S\ref{subsec:ahonestandbcheat}. Finally, in \S\ref{subsec:acheatandbcheat}, we examine the scenario $\wp^\text{AB}$ where both~A and~B cheat. 
\subsection{\texorpdfstring{$\wp^\text{0}$}{p0 }}
\label{subsec:honestqtp}
Here we present two results for~$\wp^\text{0}$,
with this QTP shown in Fig.~\ref{fig:pr0}.
We begin by considering one qubit, i.e., $m=1$.
Then we generalise our result to the case~$m>1$ qubits,
which includes a trivial case, with~C applying~$\mathds{1}^m$,
and a non-trivial case
with~C applying a non-trivial operation.
For the~$\wp^\text{0}$ protocol, A, B, and~C are required to obey the original QTP.

Here we calculate the state after each cycle for~$\wp^0$ as depicted in Fig.~\ref{fig:pr0}
for $m=1$,
which we call the isolated-qubit case as C's qubit is not entangled with any other qubits at the outset.
First C~applies the rotation gate to his input qubit~$\ket0$
yielding output qubit~(\ref{eq:qubit})
\begin{equation}
\label{eq:honestqubitc1}
\ket\Psi
\gets\sum_{i\in\{0,1\}}\psi_i\ket{i},
\end{equation}
using notation~$\ket\Psi$
for the target state as in the fidelity-threshold expression~(\ref{eq:thresholdfidelity}),
which is sent to~A.
In parallel,
D applies the gadget~(\ref{eq:entgadget}) to~$\ket{00}$
yielding~$\ket{\widetilde{00}}$
and then sends the first share to~A,
who thus receives two qubits.
After~A applies CNOT to her two qubits,
the resultant three-qubit state is
\begin{equation}
\label{eq:AandDp0}
\psi_0\ket{0\widetilde{00}}
+\psi_1\ket{1\widetilde{01}}.
\end{equation}
After~A applies H to the second qubit~(\ref{eq:AandDp0}),
she obtains
\begin{align}
&\frac12\left[\ket{00}\left(\psi_0\ket0+\psi_1\ket1\right)+\ket{01}\left(\psi_0\ket1+\psi_1\ket0\right)\right.\nonumber \\ &\left.+\ket{10}\left(\psi_0\ket0-\psi_1\ket1\right)+\ket{11}\left(\psi_0\ket1-\psi_1\ket0\right)\right].   
\end{align}
In the next cycle, A~measures both her qubits to obtain
\begin{equation}
\label{eq:P0}
   \frac12\left(\psi_0\ket{b}+(-1)^a \psi_{1}\ket{1\oplus b}\right).
\end{equation}
 Then~B applies $Z^a X^b$~(\ref{eq:B}) to this state~(\ref{eq:P0}}) to obtain
\begin{equation}
\label{eq:honestqubitb1}
\rho_{ab}^0\gets\left[\frac12(\psi_0\ket0+ \psi_1\ket1)\right]
\left[\frac12(\psi_0\ket0+ \psi_1\ket1)\right]^\dagger
\end{equation}
which implies
\begin{equation}
\label{eq:tracep0}
\tr \rho_{ab}^0 =\frac14,
\end{equation}
is independent of the pair~$(a,b)$,
using the notation for B's state in the fidelity-threshold expression~(\ref{eq:thresholdfidelity}). Trace being a $\nicefrac14$ is consistent with expectation~(\ref{eq:tracerhoab}).
As the output state~(\ref{eq:honestqubitb1})
is identical to the input state~(\ref{eq:honestqubitc1}),
\begin{equation}
\label{eq:}
f^0_\text{th}=1,\,
m=1,
\end{equation}
meaning that the threshold fidelity to prove honesty with ideal tools is unity.
Now we proceed to larger~$m$ cases.

Now we generalise $\wp^0$ to the case of an arbitrary number~$m$ of qubits
but with~C restricted to apply the trivial rotation~(\ref{eq:multiqubitrotation})
to her~$m$ qubits.
Initially C~applies the gate $\mathds{1}^m$ to obtain \begin{equation}
\label{eq:target0m}
\ket\Psi\gets\ket0^m,
\end{equation}
which is the target state~(\ref{eq:thresholdfidelity}).
Then D~applies the entanglement gadget~(\ref{eq:entgadget})
and~A applies the entanglement-gadget inverse~(\ref{eq:inverseentgadget}) to the $m^{\text{th}}$ and $(m+1)^{\text{th}}$ qubits,
yielding the resultant state
\begin{equation}
    \frac12\left(\ket0^{m-1}\otimes(\ket{000}+\ket{100}+\ket{011}+\ket{111})\right).
\end{equation}
A~measures the pair $(a,b)$ for the~$m^{\text{th}}$ and~$(m+1)^{\text{th}}$ qubits
to yield
\begin{equation}
\label{eq:P0m1}
\frac12\ket0^{m-1}\otimes\ket{b},
\end{equation}
which is independent of~$a$.
Then~B applies $Z^a X^b$ to the~$m^{\text{th}}$ qubit state~(\ref{eq:P0m1}) depending on A's measurement pair $(a,b)$ to obtain
\begin{equation}
\label{eq:rhoab0m1}
\rho_{ab}^0 \gets \frac14\ket0^m\!\bra{0} \implies f^0_{\text{th}}=1,
\end{equation}
where~$\rho_{ab}$
is independent of~$(a,b)$ as was the case for $m=1$ above and the trace agrees with the result in Eq.~(\ref{eq:tracep0}).

Next we deal with the non-trivial case where C~applies a rotation gate~(\ref{eq:entop}) that yields target state
\begin{equation}
\label{eq:Psim}
\ket{\Psi(\theta)}\gets\cos{\nicefrac{\theta}{2}}\ket0^m+ \sin{\nicefrac{\theta}{2}}\ket1^m,
\end{equation}
which is normalised to unity as it should be.
In the next cycle, D~applies the gadget~(\ref{eq:entgadget}), and~A applies the gadget~(\ref{eq:inverseentgadget}). The resultant state of the circuit is
 \begin{align}
\frac12&\left(\cos{\nicefrac{\theta}{2}}\ket0^{m-1}\otimes(\ket{000}+\ket{100}+\ket{011}+\ket{111})\right.\nonumber\\
&\left.+\sin{\nicefrac{\theta}2}\ket0^{m-1}\otimes(\ket{010}-\ket{110}+\ket{001}-\ket{101})\right).
\end{align}
A~measures the pair $(a,b)$ for the~$m^{\text{th}}$ and~$(m+1)^{\text{th}}$ qubits yielding the resultant state
\begin{equation}
\label{eq:P0m}
\frac12\left(
\cos \nicefrac\theta2\ket0^{m-1}\ket b+ (-1)^a\sin \nicefrac\theta2\ket1^{m-1}\ket{b\oplus1}
\right).
\end{equation}
Then honest B's applies $Z^a X^b$~(\ref{eq:B}) to the~$m^{\text{th}}$ qubit of the state~(\ref{eq:P0m})
to obtain
\begin{align}
\label{eq:P0m2}
\rho_{ab}^0\gets&\left[\frac12\left(
\cos\theta\ket0^{m}+\sin \theta\ket1^{m}
\right)\right] \nonumber\\
&\otimes\left[\frac12\left(
\cos\theta\ket0^{m}+\sin \theta\ket1^{m}
\right)\right]^{\dagger},
\end{align}
whose trace agrees with Eq.~(\ref{eq:tracep0}),
which implies
\begin{align}
\label{eq:fth0}
 f^0_{\text{th}}=1
\end{align}
for $\wp^0$.
\subsection{\texorpdfstring{$\wp^\text{A}$}{pA} }
\label{subsec:acheatbhonest}
Now we proceed to present our results for both cheating~A cases $\wp^{\text{A}1}$ and $\wp^{\text{A}2}$, which are depicted in Figs.~\ref{fig:prA}(a, b), respectively.
As in \S\ref{subsec:honestqtp},
we present our results for the case that $m=1$.
Then we generalise the calculation to an arbitrary number $m>1$ qubits.
For both $\wp^{\text{A}1}$ and $\wp^{\text{A}2}$ D, B~and~C are required to obey the original QTP,
but~A cheats.
\subsubsection{\texorpdfstring{$\wp^\text{A}$}{pA } for \texorpdfstring{$m=1$}{m=1}}
Here we show the results for both $\wp^{\text{A}1}$ and $\wp^{\text{A}2}$ for~$m=1$.
First we calculate the state after each cycle.
Then we calculate the fidelity threshold between the state C~sends to~A and the final state received by~C from~B.

For~$\wp^{\text{A}1}$,
A receives the target qubit~(\ref{eq:honestqubitc1}) from~C in the first cycle.
In the second cycle, D~applies the entanglement gadget~(\ref{eq:entgadget});
then the resultant state is
\begin{equation}
\label{eq:qubitA}
    \frac{\psi_0}{\sqrt{2}}\left(\ket{000}+\ket{011}\right)+\frac{\psi_1}{\sqrt{2}}\left(\ket{100}+\ket{111}\right).
\end{equation}
Next A~measures the first qubit of the three-qubit state~(\ref{eq:qubitA}) to generate
\begin{equation}
\label{eq:psimeasureA}
\left(a,
\frac12\left( \mathds1^2+\ket{00}\bra{11}+\ket{11}\bra{00}\right)\right)
\end{equation}
with the first qubit destroyed and the resultant quantum state being for the second and third qubits;
the probability of this outcome is $\abs{\psi_a}^2$.
Then~A trashes the second qubit to generate a random bit~$b$, the resultant state 
\begin{equation}
    \label{eq:rhoabA1}
    \left( a,b, \mathds1\right),
\end{equation}
with probability\begin{equation}
\label{eq:probA}
    p_{ab}^{\text{A1}} =\frac{\abs{\psi_a}^2}4.
\end{equation}
A sends the pair~$(a,b)$ to~B.
B's state,
after applying $X^bZ^a$,
is 
\begin{equation}
\label{eq:PA1B}
   \rho_{ab}^{\text{A1}}\gets \frac{\abs{\psi_a}^2}4 \mathds1 \; \forall a,b,
\end{equation}
which implies that 
\begin{equation}
    \label{eq:tracepa1}
    0\leq\tr \rho_{ab}^{\text{A1}} =\frac{\abs{\psi_a}^2}2\leq\frac12.
\end{equation}
Thus,
\begin{equation}
\label{eq:A1overlap}
\bra\Psi\rho_{ab}^{\text{A1}}\ket\Psi =\frac{\abs{\psi_a}^2}4\bra\Psi\mathds1\ket\Psi=\frac{\abs{\psi_a}^2}4= p_{ab}^{\text{A1}},
\end{equation}
which implies~(\ref{eq:thresholdfidelity})
\begin{equation}
    \label{eq:fthA11}
f^{\text{A}1}_{\text{th}}=\sum_{a,b} p_{ab}^{\text{A1}}=\frac12.
\end{equation}
Therefore, verifying that the threshold fidelity is greater than half allows the issuance of Certificate~\ref{cert:pA}.

Now we determine the fidelity threshold for $\wp^{\text{A}2}$.
In the first cycle, A~receives the qubit~(\ref{eq:honestqubitc1}) from~C.
D~applies the entanglement gadget~(\ref{eq:entgadget}) in the second cycle.
Next A~trashes the qubit~$\ket{\psi}^0$ sent from~C. Instead~A generates a two bits, namely~$(a,b)$, and sends those bits to~B.
Regardless of what~B does, his share is traced over A's share of the ebit,
yielding 
\begin{equation}
\label{eq:PA2B}  \rho^{\text{A}2}_{ab}\gets\frac{\mathds1}{8} \implies f^{\text{A}2}_{\text{th}}
=\frac12,
\end{equation}
accounting for the probability of each pair~$(a,b)$ being~$\nicefrac14$.
\subsubsection{\texorpdfstring{$\wp^\text{A}$}{pA} for \texorpdfstring{$m$}{m }~qubits~\texorpdfstring{$m>1$}{m>1 }}
Now we generalise $\wp^{\text{A}1}$ and $\wp^{\text{A}2}$ to the case of an arbitrary~$m$ of qubits. 
We start with the case $\wp^{{\text A}1}$ for which C~applies the trivial rotation gate~(\ref{eq:multiqubitrotation}) to his $m$~qubits.
A~receives the target qubit~(\ref{eq:target0m}) from~C in the first cycle.
In the second cycle D~applies the entanglement gadget~(\ref{eq:entgadget}) to the $(m+1)^{\text{th}}$ and $(m+2)^{\text{th}}$ qubits; then the resultant state is
\begin{equation}
\label{eq:rho1Am0}
\frac1{\sqrt2} (\ket{0}^{m+2}+\ket{0}^m\otimes\ket{11}).
\end{equation}
Next A~destructively measures the $m^{\text{th}}$ qubit~(\ref{eq:rho1Am0}) to obtain a bit value~$a$.
Then A~trashes the~$(m+1)^{\text{th}}$ qubit and instead generates a random bit~$b$.
A~sends the pair~$(a,b)$ to~B. Regardless of what~B does, his share is traced over A's share of the ebit, yielding 

\begin{equation}
\label{eq:PA1}
  \rho_{ab}^{\text{A1}} \gets \ket0\!^{(m-1)}\!\bra0\otimes \frac{\mathds1}4 \implies f^{\text{A}1}_{\text{th}}=\frac12,
\end{equation}
which is consistent with the threshold fidelity for $m=1$~(\ref{eq:fthA11}).

Now we proceed to the non-trivial case with C~applying the rotation gate~$R(\Psi)$~(\ref{eq:entop}) to obtain the target state~(\ref{eq:Psim}).
In the next cycle, D~applies the entanglement gadget~(\ref{eq:entgadget}). The resultant state is
\begin{align}
\frac{\cos{\nicefrac\theta2}}{\sqrt{2}}&\left(\ket0^{m+2} +\ket{0}^m\otimes \ket{11}\right)\nonumber
    \\ &+ \frac{\sin{\nicefrac\theta2}}{\sqrt{2}} \left( \ket1^{m+2} +\ket1^m\otimes \ket{00}\right).
\end{align}
In the third cycle, A~destructively measures the $m^{\text{th}}$ qubit to obtain bit value~$a$. A~trashes the $(m+1)^{\text{\text{th}}}$ qubit and generates a random bit~$b$~instead.

Regardless of what~B does, his share is traced over A's share of the ebit.
The resultant state is
\begin{align}
\label{eq:PA1m}
\rho_{ab}^{\text{A1}}(\theta)
\gets&\big(\cos^2\theta ~\ket a\!^{(m-1)}\!\bra a\nonumber\\
&+\sin^2\theta~ \ket{a\oplus1}\!^{(m-1)}\!\bra{a\oplus1}\big)\otimes \frac{\mathds1}4,
\end{align}
which implies that 
\begin{equation}
    \tr \rho_{ab}^{\text{A1}} =\frac12,
\end{equation} 
which agrees with the upper bound of Eq.~(\ref{eq:tracepa1}),
and
\begin{equation}
\label{eq:pA1fth}
\expval{\rho_{ab}^{\text{A1}}(\theta)}
{\Psi(\theta)}
\implies \frac14\left(\cos^{a\oplus 1}\nicefrac{\theta}{2}\,
\sin^a\nicefrac{\theta}2\right)^4,
\end{equation}
which implies that
\begin{equation}
 \label{eq:ftha1}
f^{\text{A}1}_{\text{th}}(\theta)=\frac12 -\frac14 \sin^2\theta.
\end{equation}
Suppose~C affects the rotation with~$\theta$ chosen uniformly randomly from the interval~$[0,\pi)$.
Then the average threshold fidelity is 
\begin{equation}
\label{eq:averagefA}
\bar{f}^{\text{A}1}_{\text{th}}
=\int^\pi_0 \frac{\dd\theta}{\pi}\left( \frac12 -\frac14 \sin^2\theta\right)=\frac38,
\end{equation}
which is the arithmetic mean of $f^{\text{A}1}_{\text{th}}(\theta)$~(\ref{eq:PA1m}) for $\theta=0$ and $\theta =\nicefrac{\pi}{2}$.

Now we proceed to present our results for  $\wp^{\text~A 2}$.
In the first cycle C~applies $\mathds1$ to obtain the target state~(\ref{eq:Psim}). 
The state of the circuit after~D applies the entanglement gadget~(\ref{eq:entgadget}) to the $(m+1)^{\text{th}}$ and $(m+2)^{\text{th}}$ qubits is 
\begin{equation}
\nicefrac1{\sqrt2} (\ket{0}^{m+2}+\ket{0}^m\otimes\ket{11}).
\end{equation}
In the next cycle, A~trashes the $m^{\text{th}}$ and $(m+1)^{\text{th}}$ qubits and instead generates a random pair~$(a,b)$
A sends the pair~$(a,b)$ to~B.
B's state after applying $X^bZ^a$ is 
\begin{equation}
\label{eq:PA2}
   \rho_{ab}^{\text{A2}}\gets\ket0\!^{(m-1)}\!\bra0\otimes \frac{\mathds1}4 \implies f^{\text{A}2}_{\text{th}}=\frac12,
\end{equation}
which agrees with $f^{\text{A}2}_{\text{th}}(0)$ because $\theta=0$ corresponds to the trivial case.

Now we proceed to the non-trivial case for $\wp^{\text{A}2}$ with C~applying the rotation gate~$R(\Psi)$~(\ref{eq:entop}) to obtain the target state~(\ref{eq:Psim}).
In the next cycle, D~applies the entanglement gadget~(\ref{eq:entgadget}) to the $(m+1)^{\text{th}}$ and $(m+2)^{\text{th}}$ qubits, yielding
\begin{align}
    \frac{\cos{\nicefrac\theta2}}{\sqrt{2}} &\left(\ket0^{m+2} +\ket{0}^m\otimes \ket{11}\right)\nonumber
    \\ &+ \frac{\sin{\nicefrac\theta2}}{\sqrt{2}} \left( \ket1^{m+2} +\ket1^m\otimes \ket{00}\right).
\end{align}
After A~trashes the $m^{\text{th}}$ and $(m+1)^{\text{th}}$ qubits and instead generates a random bit~$b$.
A~sends the pair~$(a,b)$ to~B.

Regardless of what~B does, his share is traced over A's share of the ebit, yielding 
is
\begin{align}
\label{eq:PA2m}
\rho_{ab}^{\text{A2}}(\theta)\gets&\left(\cos^2\theta~\ket0\!^{(m-1)}\!\bra0 \right.\nonumber\\  &\left. +\sin^2\theta~\ket1\!^{(m-1)}\!\bra1\right)\otimes \frac{\mathds1}8\nonumber\\
\implies& \bra{\Psi(\theta)}\rho^{\text{A}2}_{ab}(\theta)\ket{\Psi(\theta)}=\frac18\left(1-\frac12 \sin^2\theta\right)\nonumber\\
 \implies& f^{\text{A}2}_{\text{th}}(\theta) =\frac12 -\frac14 \sin^2\theta,
\end{align}
which agrees with $f^{\text{A}1}_{\text{th}}(\theta)$~(\ref{eq:ftha1}).
Suppose~C affects the rotation with~$\theta$ chosen uniformly randomly from the interval~$[0,\pi)$.
Then the average threshold fidelity is 
\begin{equation}
\label{eq:averagefA2m}
    \bar{f}^{\text{A}2}_{\text{th}}=\int^\pi_0 \frac{\dd\theta}{\pi}\left( \frac12 -\frac14 \sin^2\theta\right)=\frac38,
\end{equation}
which agrees with $\bar{f}^{\text{A}1}_{\text{th}}$~(\ref{eq:averagefA}).
\subsection{\texorpdfstring{$\wp^\text{B}$}{pB}}
\label{subsec:ahonestandbcheat}
Here we present our result for~$\wp^\text{B}$,
which is depicted in Fig.~\ref{fig:prB}.
We begin with the case of $m=1$.
Then we generalise our result to an arbitrary number~$m>1$ qubits.
In this scenario B~cheats.

Here we calculate the state after each cycle for~$\wp^\text{B}$.
First C~applies the rotation gate~(\ref{eq:bloch}) and sends the output to~A.
C's output qubit is
\begin{equation}
\label{eq:bloch1}
\ket{\Psi(\theta,\phi)} \gets\cos \nicefrac{\theta}2\ket0+\text{e}^{\text{i} \phi} \sin\nicefrac{\theta}2\ket1,
\end{equation}
where we employ complex-valued  coefficients rather than real-valued~(\ref{eq:honestqubitc1}). 
In the next cycle, D~applies the entanglement gadget~(\ref{eq:entgadget}) to the second and the third qubits to obtain
\begin{equation}
    \ket{\Psi(\theta,\phi)}\otimes\ket{\widetilde{00}}.
\end{equation}
In the second cycle, A~applies the entanglement-gadget inverse~(\ref{eq:inverseentgadget}) to the first and the second qubits to obtain
\begin{align}
\frac{\cos{\nicefrac{\theta}{2}}}{2}&\left( \ket{000}
+\ket{100}+\ket{011}+ \ket{111}\right)\nonumber\\&+\frac{\text{e}^{\text{i} \phi}\sin{\nicefrac{\theta}{2}}}{2}\left( \ket{010}-\ket{110}+\ket{001}-\ket{101}\right),
\end{align}
which has unit norm.

In the third cycle, A~destructively measures the pair$~(a,b)$ for the first and the second qubits, respectively.
In the next cycle~B trashes the qubit that D~sends him and regenerates the qubit $\ket{0}$.
Then~B applies $\mathds1^b X^a$ to his qubit to obtain
\begin{equation}
\label{eq:fabB}
\rho_{ab}^{\text{B}}\gets \frac14\ket{a} \bra{a}
\implies \tr \rho_{ab}^{\text{B}} =\frac14. 
\end{equation}
Thus,
\begin{equation}
\label{eq:fthB1}
\expval{\rho_{ab}^{\text{B}}}{\Psi(\theta,\phi)}
\gets\frac18\left(\cos^2\nicefrac{\theta}{2}~\delta_{0a} + \sin^2\nicefrac{\theta}{2}~\delta_{1a}\right),
\end{equation}
which implies that
\begin{equation}
\label{eq:fthBtheta}
f^{\text{B}}_{\text{th}}(\theta)=\nicefrac12,
\end{equation}
which depends on~$\theta$ but not on~$\phi$.

The certificate for teleportation should not depend on the choice of input state,
so the threshold fidelity~(\ref{eq:fthBtheta})
is not appropriate for our purpose.
Instead,
we adopt the approach of Grosshans and Grangier~\cite{GG01} to establish a threshold based on averaging over fidelities~(\ref{eq:fthB1}) to obtain
\begin{align}
\label{eq:fidelity2/3}
\Bar{f}_{ab}^{\text{B}}
:=&-\frac1{4\pi} \oiint \dd\phi \dd(\cos\theta) \abs{\expval{\rho_{ab}^{\text{B}}}{\Psi(\theta,\phi)}}^2 \nonumber\\
=& \frac14\left(\frac13 \delta_{a0}+\frac23\delta_{a1}\right),
\end{align}
for each~$a$.
If~A ignores the outcome $a=0$
and only retains the outcome $a=1$,
essentially postselecting teleportation on of her measurements results,
then this average fidelity~(\ref{eq:fidelity2/3})
is~$\nicefrac23$,
which is
the same average-fidelity threshold arising for the Grosshans and Grangier condition
shown as the second threshold fidelity in Eq.~(\ref{eq:fthhalftwothirds}).
Importantly, the average fidelity~(\ref{eq:fidelity2/3})
is an average fidelity but not a threshold per se,
so this effective threshold condition is actually based on custom rather than a rigorous adversarial model.

Now we generalise $\wp^{\text{B}}$ to case of an arbitrary number~$m>1$ qubits.
In the first C~applies the gate~$\mathds1$ and sends the state to~A.
A~receives the target qubit~(\ref{eq:target0m}) from~C.
For the next cycle, D~applies the entanglement gadget~(\ref{eq:entgadget}) to obtain
\begin{equation}
  \frac12 ( \ket0^{m+2} +\ket0^m \otimes \ket{11}).
\end{equation}
In the next cycle, A~applies the entanglement-gadget inverse to the $m^{\text{th}}$ and $(m+1)^{\text{th}}$ qubits.
Then A~measures the pair $(a,b)$ for the $m^{\text{th}}$ and $(m+1)^{\text{th}}$ qubits, yielding
\begin{equation}
   \rho_{ab}^{\text{B}} \gets \frac14 \ket0^{m-1} \otimes \ket a \otimes ^{m-1}\!\bra0 \otimes \bra a \implies f^{\text{B}}_{\text{th}}=\frac12,
\end{equation}
which agrees with $f^{\text{B}}_{\text{th}}(0)$~(\ref{eq:fthB1}) because $\theta=0$ corresponds to the trivial case.

Now we proceed to the non-trivial case.
In the first cycle, C~applies the rotation gate~(\ref{eq:entop}) to obtain the target state~(\ref{eq:Psim}).
In the next cycle, D~applies the entanglement gadget~(\ref{eq:entgadget}) to the $(m+1)^{\text{th}}$ and $(m+2)^{\text{th}}$ qubits; the resultant state is
\begin{align}
\label{eq:stateB}
\frac{\cos{\nicefrac\theta2}}{\sqrt{2}} &\left(\ket0^{m+2} +\ket{0}^m\otimes \ket{11}\right)\nonumber
    \\ &+ \frac{\sin{\nicefrac\theta2}}{\sqrt{2}} \left( \ket1^{m+2} +\ket1^m\otimes \ket{00}\right),
\end{align}
which has unit norm.

In the third cycle, B~trashes his qubit and replaces the qubit with $\ket 0$. Then A~applies the entanglement-gadget inverse~(\ref{eq:inverseentgadget}) to the $m^{\text{th}}$ and the $(m+1)^{\text{th}}$ qubits. Then A~destructively measures the $m^{\text{th}}$ and $(m+1)^{\text{th}}$ qubits of the state~(\ref{eq:stateB}) to obtain the state 
\begin{equation}
\label{eq:pb-Ameasurement}
\frac12\left(\cos{\nicefrac\theta2} \ket{0}^{m-1} \ket 0 +(-1)^a \sin{\nicefrac\theta2} \ket{1}^{m-1} \ket{0} \right)
\end{equation} 
for the output pair~$(a,b)$,
with a norm $\nicefrac12$ for each state~(\ref{eq:pb-Ameasurement}). 
In the next cycle, B~applies $\mathds1^b X^a$ to the $m^{\text{th}}$ qubit~(\ref{eq:pb-Ameasurement}) to obtain
\begin{align}
\label{eq:PB}
   \rho_{ab}^{\text{B}}(\theta) \gets& \frac18\left( \cos^2{\nicefrac\theta 2}\ket{0}\!^{m-1}\!\ket a~\!^{m-1}\bra{0}\bra a\right. \nonumber\\&+ (-1)^a\cos{\nicefrac\theta 2}\sin{\nicefrac\theta 2} \ket{0}^{m-1} \ket a ~ ^{m-1}\!\bra{1} \bra{a}\nonumber\\
    &+(-1)^a\cos{\nicefrac\theta 2}\sin{\nicefrac\theta 2}\ket{1}^{m-1} \ket{a} ~^{m-1}\!\bra{0} \bra a \nonumber\\
    &\left.+ \sin^2{\nicefrac\theta 2}\ket{1}^{m-1} \ket{a}~ ^{m-1}\!\bra{1} \bra{a} \right),
\end{align}
with
\begin{equation}
    \label{eq:tracepbm}
    \tr \rho_{ab}^{\text{B}}=\frac18.
\end{equation}
Thus,
\begin{align}
\label{eq:fthB}
\bra{\Psi(\theta)}\rho_{ab}^{\text{B}}(\theta) \ket{\Psi(\theta)}
\gets&\frac18
\left(\cos^a{\nicefrac{\theta}{2}}(\sin{\nicefrac{\theta}{2}})^{a\oplus1}\right)^4 \nonumber\\
     \implies &f^{\text{B}}_{\text{th}}(\theta)=\frac14 -\frac18 \sin^2\theta.
\end{align}
Suppose~C effects the rotation with~$\theta$ chosen uniformly randomly from the interval~$[0,\pi)$.
Then the average threshold fidelity is
\begin{equation}
\label{eq:averagefBm}
    \bar{f}^{~\text{B}}_{\text{th}}=\int^\pi_0 \frac{\dd\theta}{\pi}\left( \frac14 -\frac18 \sin^2\theta\right)=\frac3{16},
\end{equation}
which is half the average threshold fidelity~(\ref{eq:averagefA}) for~$\wp^{A2}$.

\subsection{\texorpdfstring{$\wp^\text{AB}$}{pAB } }
\label{subsec:acheatandbcheat}
Here we present our results for~$\wp^\text{AB}$,
which is depicted in Fig.~\ref{fig:prAB}.
We start with the case of $m=1$.
Then we generalise our result to the case of an arbitrary number~$m>1$ of qubits.
In this protocol $\wp^\text{AB}$, A~and~B both collaborate to cheat.

Here we calculate the state after each cycle for~$\wp^\text{AB}$ as depicted in Fig.~\ref{fig:prB}. Then we calculate the average fidelity between the final state and the state~C sends to~A.
First C~applies gate~(\ref{eq:bloch}) and sends the target state~ (\ref{eq:bloch1}) to~A.
In the next cycle, D~applies the entanglement gadget~(\ref{eq:entgadget}) to the second and the third qubits to obtain
\begin{equation} 
\ket{\Psi(\theta,\phi)}\otimes\ket{\widetilde{00}}.
\end{equation}
Next A~destructively measures the first qubit to obtain bit value~$a$ and trashes the second qubit.
In the next cycle~B trashes the qubit that D~sends him and regenerates the qubit $\ket{0}$. Then~B applies $X^a$ to his qubit to obtain 
\begin{equation}
    \rho_{a}^{\text{AB}}\gets \frac12\ket{a} \bra{a} \implies \tr \rho_{ab}^{\text{AB}} =\frac12.
\end{equation}
Thus,
\begin{align}
\label{eq:fthAB1}
    \expval{\rho_{ab}^{\text{AB}}}{\Psi(\theta,\phi)}\gets& \frac14\left(\cos^2\nicefrac{\theta}{2}~\delta_{0a} + \sin^2\nicefrac{\theta}{2}~\delta_{1a}\right) \nonumber\\
    \implies& f^{\text{AB}}_{\text{th}}(\theta)=\frac12.
\end{align}
Note that the threshold fidelity depends on the choices~$\theta$ and~$\phi$,
but we can average over these quantities to obtain average fidelity
\begin{align}
\label{eq:fidelity2/3AB}
\Bar{f}_{a}^{\text{AB}}
=&-\frac1{8\pi} \oiint \dd\phi \dd(\cos\theta) \abs{\expval{\rho_{ab}^{\text{AB}}}{\Psi(\theta,\phi)}}^2 \nonumber\\
=& \frac12\left(\frac13 \delta_{a0}+\frac23\delta_{a1}\right),
\end{align}
which agrees with the average fidelity~(\ref{eq:fidelity2/3}).

Now we proceed to an arbitrary number~$m>1$ qubits.
C~applies the gate~$\mathds{1}$ to the target qubit~(\ref{eq:target0m}).
Subsequently,~C sends the target state to~A.
In the next cycle, D~applies the entanglement gadget~(\ref{eq:entgadget}) to obtain
\begin{equation}
\ket0^m \otimes \ket{\widetilde{00}}.
\end{equation}
In the next cycle, A~measures $a$ in the $m^{\text{th}}$ qubit and trashes the $(m+1)^{\text{th}}$ qubits.
In the same cycle, B~trashes the $(m+2)^{\text{th}}$ qubit and applies~$X^a$ to the state~$\ket 0$.
The resultant state
\begin{equation}
\rho_a^{\text{AB}} \gets \frac14 \ket0^{m}\!\bra0 \implies f^{\text{B}}_{\text{th}}=\frac12,
\end{equation}
which agrees with $f^{\text{B}}_{\text{th}}(0)$~(\ref{eq:fthAB1}).

Now let us proceed to the non-trivial case.
In the first cycle C~applies the rotation gate~(\ref{eq:entop}) to obtain the target state~(\ref{eq:Psim}).
In the next cycle,
D~applies the entanglement gadget~(\ref{eq:entgadget}) to the $(m+1)^{\text{th}}$ and $(m+2)^{\text{th}}$ qubits; the resultant state
\begin{align}
\frac{\cos{\nicefrac\theta2}}{\sqrt{2}}&\left(\ket0^{m+2} +\ket{0}^m\otimes \ket{11}\right)\nonumber
    \\ &+ \frac{\sin{\nicefrac\theta2}}{\sqrt{2}} \left( \ket1^{m+2} +\ket1^m\otimes \ket{00}\right)
\end{align}
has unit norm.

In the next cycle, A~destructively measures the $m^{\text{th}}$ qubit to obtain the bit value~$a$ and trashes the $(m+1)^{\text{th}}$ qubit.
In the same cycle, B~trashes the $(m+2)^{\text{th}}$ qubit and applies~$X^a$ to the state~$\ket 0$.
The resultant state is
\begin{equation}
    \label{eq:rhoAB}
    \rho_a^{\text{AB}}(\theta)\gets\frac14(\cos\nicefrac{\theta}{2})^{2(a\oplus 1)}\sin^{2a}\nicefrac{\theta}{2} \ket a^m\!\bra a
    \end{equation}
with
\begin{equation}
    \label{eq:trrhoaAB}
    \tr \rho_{a}^{\text{AB}}=\frac14.
\end{equation}
Thus,
\begin{align}
\label{eq:fthAB}
    \expval{\rho_{a}^{\text{AB}}(\theta)}{\Psi(\theta)} \gets&\frac14\left((\cos\nicefrac{\theta}{2})^{a\oplus 1}\sin^a\nicefrac\theta 2\right)^4\nonumber\\
     \implies& f^{\text{AB}}_{\text{th}}(\theta)=\frac12 -\frac14 \sin^2{\theta}.
\end{align}
Suppose~C effects the rotation with~$\theta$ chosen uniformly randomly from the interval~$[0,\pi)$.
Then the average threshold fidelity is 
\begin{equation}
\label{eq:averagefABm}
    \bar{f}^{\text{AB}}_{\text{th}}=\int^\pi_0 \frac{\dd\theta}{\pi}\left( \frac12 -\frac14 \sin^2{\theta}\right)=\frac38.
\end{equation}
Now we proceed to discuss our results.
\section{Discussion} 
\label{sec:discussion}
Here we discuss our fidelity threshold results for each of the QTPs.
First, we discuss our results for $\wp^\text{0}$ QTP.
Then, we discuss our results for $\wp^{\text{A}1}$ and $\wp^{\text{A}2}$.
Next, we present our discussion of results for $\wp^\text{B}$ QTP.
Finally, we discuss the results for $\wp^\text{AB}$ QTP.
\paragraph{$\wp^\text{0}$ QTP:} Here we discuss our results for $\wp^\text{0}$ QTP.
We obtain $f_{ab}=\nicefrac14$ for all four measurement outcome pairs $(a,b)$, regardless of the state being teleported. 
The fact that $f_{ab}$ is the same for all pairs $(a,b)$ is the consequence of~A and~B knowing nothing about the state.
The threshold fidelity for proving ideal QTP is unity because the resultant state equals to the original state in this perfect case.
Note that this argument holds for both $m=1$ and $m>1$.
For didactic purposes,
we introduce Certificate~\ref{cert:p0} for this ideal case,
while noting that no real scheme would ever achieve the performance required to obtain that certificate.
\paragraph{ $\wp^{\text{A}1}$ and $\wp^{\text{A}2}$ QTPs:} 
Next we discuss our results for both 
$\wp^{\text{A}1}$ and $\wp^{\text{A}2}$,
corresponding to~A cheating by trashing one or both qubits,
respectively.
A~can cheat in two ways:
first,
A~trashes her qubit from~D, 
replaces with a random bit,
and measures her qubit from~C;
second,
A~trashes both of her qubits from~D and~C and replaces them with two random bits.
A~sends the two bits to~B, and B~reconstructs~(\ref{eq:PA1}) and (\ref{eq:PA2}) for $\wp^{\text{A}1}$ and $\wp^{\text{A}2}$, respectively. The fidelity for state reconstruction is determined by sampling over many instances. 
We have shown that,
for the case that~C sends~A an isolated qubit ($m=1$),
B manages to reconstruct the qubit with a fidelity threshold of one half in the ideal setting.
Therefore,
Certificate~\ref{cert:pA}
is not awarded in this case
meaning that~A did not provably use any quantum resources.

Now consider the case that~C sends~A a qubit, which is entangled with qubits not involved in the teleportation,
i.e., not an isolated qubit.
We have shown that the teleportation fidelity depends on the polar angle but not on the azimuthal angle of the qubit rotation~(\ref{eq:SU2U1}).
Specifically,
we show that the fidelity is bounded above by one half in the best case,
corresponding to teleporting either~$\ket0$ or~$\ket1$,
and bounded below by one quarter in the worst case,
corresponding to teleporting a qubit that is an equal superposition of~$\ket0$ and~$\ket1$.
Interpolating between those cases, the fidelity threshold is a sinusoidal function of the polar angle 
$\theta$ of the qubit rotation.

\paragraph{$\wp^\text{B}$ QTP:} 
Now we discuss our results for $\wp^\text{B}$,
corresponding to~B cheating by trashing the qubit he receives from~D and replaces this qubit with $\ket{0}$.
The resultant constructed state has some fidelity that we calculate.
We have shown that,
for the case that~C sends the same qubit over and over to~A,
B~reconstructs the qubit with a fidelity of one half.
If C~sends a qubit chosen uniformly randomly to~A, 
then the resulting average fidelity is two-thirds,
obtained by A~post-selecting on the outcome~1 of her first qubit.
Now we consider the case that~C sends~A a qubit that is entangled with qubits not involved in the quantum teleportation. 
We have shown that the quantum teleportation fidelity depends on the polar angle.
Specifically,
we showed that the fidelity is bounded above by one quarter in the best case and bounded by one eighth in the worst case.

\paragraph{ $\wp^\text{AB}$ QTP:} Finally,
we discuss our results for $\wp^\text{AB}$,
corresponding to A~trashing her qubit from~D
and measuring her qubit from~C. A~sends the outcome of the measurement to~B.
And B~is cheating by trashing his qubit from~D and replacing the qubit with $\ket{0}$.
The resultant constructed state has some fidelity that we calculate.
We have shown that,
if C~sends the same qubit over and over to~A,
B~reconstructs the qubit with a fidelity of one half.
If C~sends a qubit chosen uniformly randomly to~A, 
then the resulting average fidelity is two-thirds,
obtained by A~post-selecting on the outcome~1 of her first qubit.
Now we consider the case that~C sends~A a qubit that is entangled with qubits not involved in the quantum teleportation. 
We have shown that the quantum teleportation fidelity depends on the polar angle.
Specifically,
we showed that the fidelity is bounded above by one half in the best case and bounded by one quarter in the worst case.

\section{Conclusions} 
\label{sec:conclusions}
We have developed an operational framework for adversarial quantum teleportation to establish fidelity thresholds for certifying quantum teleportation protocols.
Our formulation of these thresholds in terms of user‑oriented certificates provides a systematic way to guarantee that genuinely quantum resources are employed by some or all parties.
Furthermore,
we connect our fidelity thresholds to those that have arisen in the quantum teleportation literature.

Our formalisation of fidelity thresholds in terms of certificates is based on adversarial models in which~A or~B or both~A and~B may cheat.
We account for~A being able to cheat in two distinct ways by trashing one or both qubits.
These adversarial models explain fidelity thresholds of $\nicefrac12$ and~$\nicefrac23$, which are the two cases that arise historically.
Whereas evaluating quantum teleportation is typically done for teleporting an isolated qubit,
in practice, teleportation would usually be evaluated for the teleported qubit being entangled with ancill\ae.
Notably,
the threshold fidelity depends on specifics of states,
represented by the polar angle of the qubit-rotation polar angle appearing in the expression for fidelity. The certificate should be blind to the specifics of state preparation, which then leads to the requirement of using average fidelity.
Thus, our approach elucidates the ubiquity of average fidelity for quantum teleportation.

Our work is satisfying in that it introduces an operational approach that clarifies the threshold fidelities appearing in historic quantum teleportation papers. 
Moreover, establishing quantum teleportation on a strong and rigorous mathematical and operational footing is important because quantum teleportation has become central to several strategies for building quantum computers~\cite{SVN25,SKD25}.
Some of these strateegies include teleportation quantum computation~\cite{jozsa06} and measurement-induced physics in quantum computing~\cite{Goo23}
and distributed quantum computing~\cite{MDN25}.
Quantum teleportation is also vital for the burgeoning fields of quantum networks~\cite{HGL23} and quantum internet~\cite{PAA24}.
In all these cases, quantum teleportation is implemented via physical devices,
and certifying the correct operation of such devices is critical for technology development.
Our certification framework provides a natural way to achieve this, elucidating how fidelity-based benchmarks can be systematically pursued.
\section*{Acknowledgements}
\label{sec:Acknowledgments} 
This project has been supported by NSERC, NAEM, and the Queen Elizabeth Scholars Program.  
We acknowledge the traditional owners of the land on which this work was undertaken at the University of Calgary: the Treaty 7 First Nations www.treaty7.org
\bibliographystyle{unsrt}
\nocite{*}
\bibliography{aqt}
\end{document}